\documentclass{LMCS}

\usepackage{ifpdf}
\usepackage{latexsym}
\usepackage{amssymb}
\usepackage{amsmath}
\usepackage{stmaryrd}
\usepackage{mathrsfs}
\usepackage{proof}
\usepackage{verbatim}
\usepackage{graphicx}
\usepackage{enumerate,hyperref}

\def\P{\mathfrak{P}}
\def\N{\mathbb{N}}
\def\M{\mathscr{M}}

\def\PAtwo{\ensuremath{\text{PA}2}}
\def\PAext{\ensuremath{\text{PA}2^{+}}}
\def\Pred{\mathrm{pred}}
\def\Neg{\mathrm{neg}}
\def\Null{\mathrm{null}}
\def\Nat{\mathrm{nat}}

\def\limp{\Rightarrow}

\def\forallN{\forall^{\mathbb{N}}}
\def\existsN{\exists^{\mathbb{N}}}
\def\conv{\cong}
\def\TYPJ#1#2#3{#1~{\vdash}_{\mathrm{NJ}}~#2~{:}~#3}
\def\TYPK#1#2#3{#1~{\vdash}_{\mathrm{NK}}~#2~{:}~#3}

\def\K{\mathcal{K}}

\def\Lam#1#2{\lambda#1\,.\,#2}
\def\Let{\mathsf{let}}
\def\LetP#1#2#3#4{\mathsf{let}~\<#1;#2\>=#3~\textsf{in}~#4}
\def\Pair{\mathsf{pair}}
\def\Fst{\mathsf{fst}}
\def\Snd{\mathsf{snd}}

\def\Zero{\mathsf{0}}
\def\Succ{\mathsf{s}}
\def\Rec{\mathsf{rec}}
\def\Stop{\mathsf{stop}}
\def\Quote{\mathsf{quote}}
\def\Print{\mathsf{print}}
\def\cc{\mathsf{c\!c}}
\def\k{\mathsf{k}}
\def\FV{\mathit{FV}}
\def\dom{\mathrm{dom}}
\def\eval{\succ}
\def\red{\succ}
\def\wred{\succ_{\!\!\mathit{w}}}
\def\ired{\succ_{\!\mathit{i}}}
\def\jred{\succ_{\!\mathit{I}}}

\def\realJ{\Vdash_{\mathrm{NJ}}}
\def\SAT{\mathbf{SAT}}
\def\Val{\mathbf{Val}}
\def\realK{\Vdash_{\mathrm{NK}}}
\def\urealK{\Vvdash_{\mathrm{NK}}}
\def\Bot{{{\bot}\mskip-11mu{\bot}}}

\def\Int#1{\llbracket#1\rrbracket}

\outer\long\def\COUIC#1{}
\let\ds\displaystyle
\def\<{\langle}
\def\>{\rangle}
\def\notR{\lnot_R}
\def\notnotR{\lnot_R\lnot_R}
\def\nn{{{\lnot}{\lnot}}}

\newenvironment{footmath}
{\begingroup\footnotesize$$}{$$\endgroup\ignorespaces}

\def\doi{7 (2:2) 2011}
\lmcsheading%
{\doi}
{1--47}
{}
{}
{Nov.~21, 2009}
{Apr.~22, 2011}
{}

\begin{document}
\title[Existential witness extraction]{Existential witness extraction in classical\\
  realizability and via a negative translation}
\author{Alexandre Miquel}
\address{LIP (UMR 5668 -- CNRS -- ENS de Lyon -- UCBL -- INRIA)
  ENS de Lyon, Universit\'e de Lyon, France}
\email{alexandre.miquel@ens-lyon.fr}

\begin{abstract}
  We show how to extract existential witnesses from classical proofs
  using Krivine's classical realizability---where classical proofs are
  interpreted as $\lambda$-terms with the call$/$cc control operator.
  We first recall the basic framework of classical realizability (in
  classical second-order arithmetic) and show how to extend it with
  primitive numerals for faster computations.
  Then we show how to perform witness extraction in this framework, by
  discussing several techniques depending on the shape of the
  existential formula.
  In particular, we show that in the $\Sigma^0_1$-case, Krivine's
  witness extraction method reduces to Friedman's through a
  well-suited negative translation to intuitionistic second-order
  arithmetic.
  Finally we discuss the advantages of using call$/$cc rather than a
  negative translation, especially from the point of view of an
  implementation.
\end{abstract}

\titlecomment{This paper is an expanded version of~\cite{Miq09}.}

\maketitle

\section{Introduction}

\noindent Extracting an existential witness (i.e.\ an object~$t$ such that
$A(t)$) from a proof of the formula $\exists x\,A(x)$ is now a
well-understood technique in intuitionistic logic.
The simplest way to do it is to normalize the proof and retrieve
the witness from the premise of its normal form.
Through the Brouwer-Heyting-Kolmogorov interpretation, one can also
read the proof as a functional program that reduces to a pair
whose first component is the desired witness.
Such techniques are implemented in proof-assistants based on
intuitionistic systems~\cite{CoqMan09,Let02,Sch05}.

Extracting a witness from a classical proof of an existential formula
is much more difficult, since classical logic is known not to enjoy
the witness property.
Such an extraction is actually not always feasible:
for instance, we cannot expect to extract a witness from the obvious
classical proof of the formula
$$\exists x\,((x=1\land C)\lor(x=0\land\lnot C))$$
in general---think of~$C$ being undecidable or, say, Riemann's
conjecture.

However, several techniques~\cite{Kre51,Kre52,God58,Fri78,Koh08} have
been proposed in order to extract a witness from a classical proof of
an existential formula in some particular cases---typically: when the
formula is~$\Sigma^0_1$ (i.e.\ of the form
$\exists x\,f(x)=0$).

\subsection{Friedman's method}

One of the most popular methods to extract witnesses from classical
proofs of $\Sigma^0_1$-formul{\ae} has been introduced by
Friedman~\cite{Fri78}.
The idea of Friedman is to generalize G{\"o}del and Kolmogorov's
double negation translation by replacing the intuitionistic negation
$\lnot A\equiv A\limp\bot$ by a relative negation
$\notR A\equiv A\limp R$ parameterized by an arbitrary formula~$R$.
(The only condition on~$R$ is that its free variables should not be
captured in the formula or the proof we want to translate.)
In first-order Peano arithmetic (PA) for instance, this negative
$R$-translation $A\mapsto A^{\nn}$ can be defined as follows
$$\begin{array}{r@{~~}c@{~~}l@{\qquad}r@{~~}c@{~~}l}
  (e_1=e_2)^{\nn} &\equiv& \notnotR(e_1=e_2) &
  (\lnot A)^{\nn} &\equiv& \notR A^{\nn} \\
  (A\land B)^{\nn} &\equiv& A^{\nn}\land B^{\nn} &
  (\forall x\,A)^{\nn} &\equiv& \forall x\,A^{\nn} \\
  (A\lor B)^{\nn} &\equiv& \notR(\notR A^{\nn}\land\notR B^{\nn}) &
  (\exists x\,A)^{\nn} &\equiv& \notR\forall x\notR A^{\nn} \\
\end{array}$$
and it is easy to check that if a formula~$A$ is provable in Peano
arithmetic, then the formula~$A^{\nn}$ is provable in Heyting
Arithmetic (HA), independently from the choice of the formula~$R$.

If we apply this translation to a classical proof~$p$ of the formula
$\exists x\,f(x)=0$ (i.e.\ a $\Sigma^0_1$-formula), then we get an
intuitionistic proof $p^*$ of the formula
$$\notR\forall x\notR\notR\notR f(x)=0\,.$$
By simplifying the triple (relative) negation and by unfolding the
relative negation $\notR A\equiv A\limp R$, we thus get an
intuitionistic proof ${p^*}'$ of the formula
$$\forall x\,(f(x)=0\limp R)~\limp~R\,.$$
(The proof~${p^*}'$ we get is parametric w.r.t. the formula~$R$.)

Now, let us introduce Friedman's trick, which is to instantiate the
parameter~$R$ with the formula we want to prove, letting
$R\equiv\exists y\,f(y)=0$.
Thus ${p^*}'$ is an intuitionistic proof of the implication
$$\forall x\,\bigl(f(x)=0~\limp~\exists y\,f(y)=0\bigr)
~\limp~\exists y\,f(y)=0\,.$$
whose left member is the introduction rule of existential
quantification.
Combining the modus ponens with the introduction rule of existential
quantification, we finally get an intuitionistic proof
$\tilde{p}\equiv({p^*}')\,(\exists\text{-intro})$ of the formula
$$\exists y\,f(y)=0$$
from which we can perform the standard extraction techniques.

The transformation above actually shows that classical arithmetic is
conservative over intuitionistic arithmetic on the class of
$\Sigma^0_1$-formul{\ae}.
Since the transformation even works when the inner formula depends on
free variables, it is easy to generalize the latter result to a result
of conservativity on the class of $\Pi^0_2$-formul{\ae}:
$$\infer[\rlap{\footnotesize($\forall$-intro)}]
{\vdash_{\text{HA}}\forall x\,\exists y\,f(x,y)=0}{
  \infer[\rlap{\footnotesize(Friedman's transformation)}]
  {\vdash_{\text{HA}}\exists y\,f(x_0,y)=0}{
    \infer[\rlap{\footnotesize($\forall$-elim, $x_0$ fresh)}]
    {\vdash_{\text{PA}}\exists y\,f(x_0,y)=0}{
      \vdash_{\text{PA}}\forall x\,\exists y\,f(x,y)=0
    }
  }
}$$
This conservativity result has been extended by Friedman (using the
same technique) to much stronger pairs of classical and
intuitionistic theories, such as PA2$/$HA2, \dots,
PA$\omega/$HA$\omega$, Z$/$IZ, ZF$/$IZF$_C$~\cite{Fri78}.

\subsection{Krivine's classical realizability}

Up to the 90's, the computational contents of classical proofs was
only studied indirectly, via clever translations to intuitionistic
logic~\cite{God58,Kre51,Fri78} or to linear logic.
% (that can be seen as a symmetric form of intuitionistic logic).
The situation quickly changed with the discovery of a strong
connection between classical reasoning principles (such as Peirce's
law) and control operators (such as call$/$cc)~\cite{Gri90}.
This led to the rise of many extensions of the $\lambda$-calculus with
control primitives, such as Krivine's
$\lambda_c$-calculus~\cite{Kri05}, Parigot's
$\lambda\mu$-calculus~\cite{Par97}, Barbanera and Berardi's (non
deterministic) symmetric $\lambda$-calculus~\cite{BB96} or Curien and
Herbelin's $\lambda\bar{\lambda}\mu\tilde{\mu}$-calculus~\cite{CH00}.
(This list is far from being exhaustive.)

Among these different proposals to extend the proofs-as-programs
paradigm to classical logic, Krivine's theory of classical
realizability~\cite{Kri01,Kri05} enjoys a particular position.
First, it is based on realizability rather than on typing, which makes
it naturally more flexible and more powerful than systems that are
simply based on typing.
Second, the simplicity on the underlying calculus of realizers
(the $\lambda$-calculus extended with the call$/$cc control
primitive) and of its evaluation policy (weak head normalization) hides
its main feature, which is its ability to incorporate new instructions
in order to realize new formul{\ae}, such as (for instance) several
forms of the axiom of choice~\cite{Kri03}.
Although classical realizability is traditionally presented in
second-order classical arithmetic, it can be extended to much more
expressive logical frameworks such as Zermelo-Fraenkel set
theory~\cite{Kri01} or the calculus of constructions with
universes~\cite{Miq07}.

Less known is the fact that Krivine's framework allows to perform
classical witness extraction directly (especially from realizers of
$\Sigma^0_1$-formul{\ae}), without going through a negative
translation such as Friedman's.
The purpose of this paper is twofold.
First, it aims at presenting some methods that naturally come with
classical realizability in order to extract witnesses from classical
proofs of existential formul{\ae}---especially
$\Sigma^0_1$-formul{\ae}.
Second, it aims to relate the extraction method for
$\Sigma^0_1$-formul{\ae} with Friedman's, by showing that through a
well-chosen negative translation (inspired from~\cite{Oli08}), both
methods are basically the same (up to the details of the translation).

One of the difficulties of tracking \emph{arithmetic} reasoning
through a negative translation is that some parts of the proof carry
over logical invariants whereas other parts are only devoted to
arithmetic computations.
To solve this problem, we shall introduce \emph{primitive numerals} in
the language of realizers, while showing that they (essentially)
realize the same formul{\ae} as Church numerals.
As a side effect, replacing Church numerals with primitive numerals
also makes the corresponding extraction technique much more
realistic---and we believe, much more efficient---in the perspective
of a practical implementation.

\subsection{Outline of the paper}

In section~\ref{s:PA2}, we present a type system for classical
second-order arithmetic (PA2) based on the $\lambda$-calculus extended
with the primitive call$/$cc.
This type system is given its semantics in section~\ref{s:RealK},
by defining a family of classical realizability models
(following~\cite{Kri05}).
In section~\ref{s:PrimInt}, we extend the calculus of realizers and
the type system for PA2 with primitive numerals to make arithmetic
computations more efficient (in proof-terms) and more easily tractable
through the negative translation.
The classical witness extraction methods are presented in
section~\ref{s:Witness} and we illustrate them with an example based
on the minimum principle in section~\ref{s:Example}.
In section~\ref{s:HA2}, we define a more traditional type system for
intuitionistic second-order arithmetic (HA2), which we relate to the
type system for PA2 by defining in section~\ref{s:NegTrans} a negative
translation in the spirit of~\cite{Oli08}.

%This paper is an expanded version of~\cite{Miq09}.

\section{Classical second-order arithmetic (\PAtwo)}
\label{s:PA2}

\subsection{The language of second-order arithmetic}
\label{ss:LangPA2}

The language of \PAtwo\ (Fig.~\ref{f:PA2} p.~\pageref{f:PA2}) is
made of two kinds of syntactic expressions:
\emph{arithmetic expressions} (a.k.a.\ first-order terms%
\footnote{We shall prefer the terminology of `arithmetic expression'
  to the more standard terminology of `first-order term' to prevent a
  confusion with the proof-terms we shall introduce in
  section~\ref{ss:TypingPA2}.})
that represent individuals, and \emph{formul{\ae}} that represent
mathematical propositions.

\begin{figure}[p]
  \def\myskip{\vskip 12pt}
  $$\begin{array}{@{}c@{}}
    \hline\hline
    \noalign{\medskip}
    \underline{\textbf{The language of PA2}} \\
    \noalign{\bigskip}
    \begin{array}{l@{}r@{\quad}r@{\quad}l}
      \textbf{Arithmetic expr.}
      & e &::=& x \quad|\quad f(e_1,\ldots,e_k) \\[6pt]
      \textbf{Formul{\ae}}
      & A,B &::=& \Null(e) \quad|\quad X(e_1,\ldots,e_k) \\
      &&|& A\limp B\quad|\quad\forall x\,A\quad|\quad\forall XA \\[6pt]
      \textbf{Proof-terms}
      & t,u &::=& x \quad|\quad \Lam{x}{t} \quad|\quad tu
      \quad|\quad \cc \\[6pt]
      \textbf{Contexts}
      & \Gamma &::=& \emptyset \quad|\quad \Gamma,x:A \\
    \end{array} \\
    \noalign{\bigskip}
    \underline{\textbf{The congruences $e\conv e'$ and $A\conv A'$}} \\
    \noalign{\bigskip}
    \begin{array}{rcl@{\quad~}rcl@{\quad~}rcl}
      0+y       &\conv& y &
      \Pred(0)  &\conv& 0 &
      \Neg(0)   &\conv& 1 \\
      s(x)+y       &\conv& s(x+y) &
      \Pred(s(x))  &\conv& x &
      \Neg(s(x))   &\conv& 0 \\
    \end{array}\quad(\text{etc.})\\
    \noalign{\medskip}
    \Null(s(x))\conv\bot \\
    \noalign{\bigskip}
    \underline{\textbf{Abbreviations}} \\
    \noalign{\bigskip}
    \begin{array}{r@{\quad}c@{\quad}l}
      \top &\equiv& \Null(0) \\
      \bot &\equiv& \forall Z\,Z \\[3pt]
      \lnot A &\equiv& A\limp\bot \\[3pt]
      A\land B &\equiv& \forall Z\,((A\limp B\limp Z)\limp Z) \\
      A\lor B &\equiv& \forall Z\,((A\limp Z)\limp(B\limp Z)\limp Z) \\[3pt]
      \exists x\,A(x) &\equiv&
      \forall Z\,(\forall x\,(A(x)\limp Z)\limp Z) \\
      \exists XA(X) &\equiv&
      \forall Z\,(\forall X(A(X)\limp Z)\limp Z) \\[3pt]
      e=e' &\equiv& \forall Z\,(Z(e)\limp Z(e')) \\
      \Nat(e) &\equiv& \forall Z\,(Z(0)\limp
      \forall y\,(Z(y)\limp Z(s(y)))\limp Z(x)) \\
    \end{array} \\
    \noalign{\bigskip}
    \underline{\textbf{Typing rules of PA2}} \\
    \noalign{\bigskip}
    \infer[\scriptstyle(x:A)\in\Gamma]{\TYPK{\Gamma}{x}{A}}{}
    \qquad\qquad
    \infer[\scriptstyle\FV(t)\subseteq\dom(\Gamma)]
    {\TYPK{\Gamma}{t}{\top}}{} \\
    \noalign{\myskip}
    \infer{\TYPK{\Gamma}{\cc}{((A\limp B)\limp A)\limp A}}{}
    \qquad\qquad
    \infer[\scriptstyle A\conv A']{\TYPK{\Gamma}{t}{A'}}{
      \TYPK{\Gamma}{t}{A}
    } \\
    \noalign{\myskip}
    \infer{\TYPK{\Gamma}{\Lam{x}{t}}{A\limp B}}{
      \TYPK{\Gamma,x:A}{t}{B}
    } \qquad\qquad
    \infer{\TYPK{\Gamma}{tu}{B}}{
      \TYPK{\Gamma}{t}{A\limp B} &\quad \TYPK{\Gamma}{u}{A}
    } \\
    \noalign{\myskip}
    \infer[\scriptstyle x\notin\FV(\Gamma)]
    {\TYPK{\Gamma}{t}{\forall x\,A}}{
      \TYPK{\Gamma}{t}{A}
    } \qquad\qquad
    \infer{\TYPK{\Gamma}{t}{A\{x:=e\}}}{
      \TYPK{\Gamma}{t}{\forall x\,A}
    } \\
    \noalign{\myskip}
    \infer[\scriptstyle X\notin\FV(\Gamma)]
    {\TYPK{\Gamma}{t}{\forall XA}}{
      \TYPK{\Gamma}{t}{A}
    } \qquad\qquad
    \infer{\TYPK{\Gamma}{t}{A\{X(x_1,\ldots,x_k):=B\}}}{
      \TYPK{\Gamma}{t}{\forall XA}
    } \\
    \noalign{\medskip}
    \hline\hline
  \end{array}$$\vspace{-12pt}
  \caption{Classical second-order arithmetic (PA2)}
  \label{f:PA2}
\end{figure}

Arithmetic expressions (notation: $e$, $e'$, $e_1$, etc.) are built
from an infinite set of first-order variables (notation: $x$, $y$,
$z$, etc.) using function symbols (notation: $f$, $g$, $h$, etc.)
defined in a given first-order signature.
Here, we assume that the signature contains a constant symbol~`$0$' for
zero, a unary function symbol~`$s$' for the successor function, and
more generally, a function symbol~$f$ of arity~$k$ for every primitive
recursive definition of a function with $k$~arguments.
In the sequel, we shall use binary function symbols `$+$' (addition)
and `$\times$' (multiplication) as well as unary function symbols
`$\Pred$' (predecessor) and `$\Neg$' (boolean negation) with the
following definitions:
$$\begin{array}{rcl@{\qquad}rcl}
  0+y &=& y & 0\times y &=& 0 \\
  s(x)+y &=& s(x+y) & s(x)\times y &=& (x\times y)+y \\[3pt]
  \Pred(0) &=& 0 & \Neg(0) &=& 1 \\
  \Pred(s(x)) &=& x & \Neg(s(x)) &=& 0 \\
\end{array}$$
(writing $1=s(0)$, $2=s(1)$, $3=s(2)$, etc.)
The set of all free variables of an arithmetic expression~$e$ is
written $\FV(e)$.
The notion of (first-order) substitution in an arithmetic expression
is defined as usual and written $e\{x:=e'\}$.

Formul{\ae} of the language of second-order arithmetic (notation: $A$,
$B$, $C$, etc.) are formed from second-order variables (notation: $X$,
$Y$, $Z$, etc.) of all arities using implication and first- and
second-order universal quantification (Fig.~\ref{f:PA2}).
We slightly deviate from the traditional presentation of the
syntax of the language~\cite{Gir89,Kri93} by explicitly introducing a
unary predicate symbol `$\Null$' expressing that its argument yields
zero.
The main reason for introducing this symbol is that it facilitates the
construction of a simple proof-term for Peano's 4th axiom within the
type system presented in section~\ref{ss:TypingPA2}.

The set of all free (first- and second-order) variables of a
formula~$A$ is written $\FV(A)$.
The notions of first- and second-order substitution in a formula are
defined as usual, and written $A\{x:=e\}$ and
$A\{X(x_1,\ldots,x_k):=B\}$ respectively.
(See~\cite{Gir89,Kri93} for a more detailed presentation of the two
forms of substitutions.)

\subsubsection{Second-order encodings}\quad
Propositional units ($\top$ and $\bot$), negation, conjunction,
disjunction, first- and second-order existential quantification as
well as Leibniz equality are represented using the second-order
encodings given in Fig.~\ref{f:PA2}.
Here, we define the propositional constant~$\top$ as a shorthand for
the formula~$\Null(0)$, which is consistent with the type system of
section~\ref{ss:TypingPA2} and the realizability interpretation of
section~\ref{s:RealK}.
Intuitively, the formula~$\top$ is the type of \emph{all} proof-terms,
and it is important not to confuse it with the (true) formula
$\mathbf{1}\equiv\forall Z\,(Z\limp Z)$ that has much less
proof-terms.

\subsection{The congruences $e\conv e'$ and $A\conv A'$}
\label{ss:KConv}

We introduce two congruences $e\conv e'$ and $A\conv A'$ over
arithmetic expressions and formul{\ae} that will be used to
incorporate the definitional equalities of the function symbols of the
signature in the conversion rule of the type system we shall introduce
in section~\ref{ss:TypingPA2}.
The same mechanism will be used to build proof-terms for Peano's 3rd
and 4th axioms.

The congruence $e\conv e'$ over arithmetic expressions is simply
defined as the congruence generated by the defining equations of
the primitive recursive function symbols of the signature.
(We already gave the equations associated with the function symbols
`$+$', `$\times$', `$\Pred$' and `$\Neg$' in
section~\ref{ss:LangPA2}.)
Of course, these equations can be oriented in such a way that they
form a confluent and terminating system of rewrite rules, so that the
congruence $e\conv e'$ is decidable.
But we shall not need such a level of detail in the sequel.

The congruence $A\conv A'$ over formul{\ae} is defined by adding the
equation $\Null(s(x))\conv\bot$ to the system of equations defining the
congruence $e\conv e'$.
Again, this new equation can be oriented from left to right so that
the resulting system of rewrite rules (including the rewrite rules for
function symbols) is confluent and terminating, and the congruence
$A\conv A'$ is thus decidable.

\subsection{A type system for classical second-order arithmetic}
\label{ss:TypingPA2}

The type$/$proof system of PA2 closely follows the spirit of
Second-order functional arithmetic (FA2)~\cite{Kri93}.
As in FA2, first- and second-order universal quantifications are
treated uniformly, by using Curry-style proof-terms that do not keep
track of introduction and elimination of universal quantifiers.%
\footnote{For this reason, a (Curry-style) proof-term should not be
  confused with the proof (i.e.\ the derivation) it comes from,
  since the latter contains much more information that cannot be
  reconstructed from the proof-term.
  In such a setting, the proof-term is merely a computational digest
  of the formal proof, where some computationally irrelevant parts of
  the proof have been already removed.}
As usual in such a framework, numeric quantifications require a
special treatment we shall recall in Section~\ref{ss:Induction}.

Formally, the type system of PA2 is based on a typing judgment of the
form $\TYPK{\Gamma}{t}{A}$, where~$\Gamma$ is a \emph{typing context},
$t$ a (Curry-style) \emph{proof-term}, and where~$A$ is a formula of
the language of PA2 (section~\ref{ss:LangPA2}).

Proof-terms of PA2 (notation: $t$, $u$, etc.) are just pure
$\lambda$-terms%
\footnote{Proof variables (i.e.\ variables of the $\lambda$-calculus)\
  are written $x$, $y$, $z$, etc.\ in the sequel, but it is important
  not to confuse them with first-order variables (written using the
  same letters) that occur in arithmetic expressions and formul{\ae}.}
enriched with a special constant~$\cc$ (`call$/$cc') to prove Peirce's
law.
The operational semantics of proof-terms (that slightly differs from
the traditional operational semantics of pure $\lambda$-calculus) will
be given in section~\ref{s:RealK}.

A typing context (notation: $\Gamma$, $\Gamma'$, $\Gamma_1$, etc.) is
a finite unordered list of declarations of the form $\Gamma\equiv
x_1:A_1,\ldots,x_n:A_n$ where $x_1,\ldots,x_n$ are pairwise distinct
proof-variables and where $A_1,\ldots,A_n$ are arbitrary formul{\ae}.
Given a typing context~$\Gamma\equiv x_1:A_1,\ldots,x_n:A_n$, we write
$\dom(\Gamma)=\{x_1;\ldots;x_n\}$ and
$\FV(\Gamma)=\FV(A_1)\cup\cdots\cup\FV(A_n)$.

The inference rules for the judgment $\TYPK{\Gamma}{t}{A}$ are given
in Fig.~\ref{f:PA2}.
These rules contain the standard typing rules of AF2~\cite{Kri93}
(that correspond to the deduction rules of intuitionistic natural
deduction in second-order predicate logic), plus a typing rule
for the constant~$\cc$ (Peirce's axiom) to recover classical logic.
These rules also contain a conversion rule as well an introduction
rule for the propositional constant~$\top$.
(These rules are specifically needed to build proof-terms for the
axioms of arithmetic.)
In particular:
\begin{enumerate}[$\bullet$]
\item For all arithmetic expressions
  $e_1(x_1,\ldots,x_k)$ and $e_2(x_1,\ldots,x_k)$ depending on the
  variables $x_1,\ldots,x_k$ such that
  $e_1(x_1,\ldots,x_k)\conv e_2(x_1,\ldots,x_k)$, we have
  $$\TYPK{}{~\Lam{z}{z}~}{~
    \forall x_1\cdots\forall x_k~
    e_1(x_1,\ldots,x_k)=e_2(x_1,\ldots,x_k)}$$
  (where $=$ stands for Leibniz equality).
  So that $\Lam{z}{z}$ is a proof-term for all definitional equalities
  attached to the function symbols of the signature.
\item Given an arbitrary proof-term~$u$ such that
  $\FV(u)\subseteq\{z\}$, we have
  $$\begin{array}{l}
    \TYPK{}{~\Lam{z}{z}\hphantom{u}~}{~
      \forall x\,\forall y\,(s(x)=s(y)\limp x=y)} \\
    \TYPK{}{~\Lam{z}{zu}~}{~\forall x\,\lnot(s(x)=0)} \\
  \end{array}$$
  so that Peano's 3rd and 4th axioms are provable in our type system.
  (The corresponding derivations are given in Fig.~\ref{f:P34Deriv}.)
\end{enumerate}

\begin{figure}[htp]
  $$\begin{array}{c}
    \hline\hline
    \noalign{\bigskip}
    \infer{\TYPK{}{\Lam{z}{z}}{
        \forall x\,\forall y\,(s(x)=s(y)\limp x=y)}}{
      \infer{\TYPK{}{\Lam{z}{z}}{\forall y\,(s(x)=s(y)\limp x=y)}}{
        \infer{\TYPK{}{\Lam{z}{z}}{s(x)=s(y)\limp x=y}}{
          \infer{\TYPK{z:s(x)=s(y)}{z}{x=y}}{
            \infer{\TYPK{z:s(x)=s(y)}{z}{
                \forall Z\,(Z(\Pred(s(x)))\limp Z(\Pred(s(y))))}}{
              \infer{\TYPK{z:s(x)=s(y)}{z}{
                  Z(\Pred(s(x)))\limp Z(\Pred(s(y)))}}{
                \infer{\TYPK{z:s(x)=s(y)}{z}{s(x)=s(y)}}{}
              }
            }
          }
        }
      }
    } \\
    \noalign{\bigskip}
    \infer{\TYPK{}{\Lam{z}{zu}}{\forall x\,(s(x)=0\limp\bot)}}{
      \infer{\TYPK{}{\Lam{z}{zu}}{s(x)=0\limp\bot}}{
        \infer{\TYPK{z:s(x)=0}{zu}{\bot}}{
          \infer{\TYPK{z:s(x)=0}{z}{\top\limp\bot}}{
            \infer{\TYPK{z:s(x)=0}{z}{\Null(\Neg(s(x)))\limp\Null(\Neg(0))}}{
              \infer{\TYPK{z:s(x)=0}{z}{s(x)=0}}{}
            }
          } &
          \infer{\TYPK{z:s(x)=0}{u}{\top}}{}
        }
      }
    } \\
    \noalign{\bigskip}
    \hline\hline
  \end{array}$$\vspace{-12pt}
  \caption{Derivations for Peano's 3rd and 4th axioms}
  \label{f:P34Deriv}
\end{figure}

\subsection{Induction}
\label{ss:Induction}

It is well known~\cite{Gir89,Kri93,Kri05} that the induction principle
$$\forall Z~\bigl(Z(0)\limp\forall y\,(Z(y)\limp Z(s(y)))\limp
\forall x\,Z(x)\bigl)$$
cannot be given a (closed) proof-term in the type system we presented
above.
The reason is that first-order quantification is interpreted
uniformly (i.e.\ as an infinitary intersection type) in our
setting, whereas universal quantification over natural numbers cannot
be interpreted uniformly, for that most proofs of $A(n)$
computationally depend on the natural number~$n$.
To circumvent this difficulty, we use a well-known trick of
second-order logic which is to relativize first-order quantifications
using the predicate
$$\Nat(x)~\equiv~\forall Z~
\bigl(Z(0)\limp\forall y\,(Z(y)\limp Z(s(y)))\limp Z(x)\bigl)$$
expressing that~$x$ belongs to the smallest set of individuals
containing zero and stable under the successor function.
With this notation, the relativized form of the induction principle
$$\forall Z~\bigl(Z(0)\limp
\forall y\,(\Nat(y)\limp Z(y)\limp Z(s(y)))\limp
\forall x\,(\Nat(x)\limp Z(x))\bigl)$$
can be given a closed proof-term in our setting.
(See~\cite{Kri93} for instance.)

More generally, we associate to every formula~$A$ a formula $A^{\Nat}$
that is obtained by relativizing all the first-order quantifications
with the predicate $\Nat$.
Formally, the formula~$A^{\Nat}$ is defined by induction of~$A$ with
the equations:
$$\begin{array}{r@{~}c@{~}l}
  (\Null(e))^{\Nat} &\equiv& \Null(e) \\
  (X(e_1,\ldots,e_k))^{\Nat} &\equiv& X(e_1,\ldots,e_k) \\
  (A\limp B)^{\Nat} &\equiv& A^{\Nat}\limp B^{\Nat} \\
  (\forall x\,A)^{\Nat} &\equiv& \forall x\,(\Nat(x)\limp A^{\Nat}) \\
  (\forall XA)^{\Nat} &\equiv& \forall X(A^{\Nat}) \\
\end{array}$$
We then easily check that
\begin{prop}
  If a closed formula~$A$ is provable in classical second-order
  arithmetic (with the unrelativized induction principle), then the
  formula $A^{\Nat}$ has a closed proof-term in the type system
  defined in Fig.~\ref{f:PA2}.
\end{prop}

\section{Classical realizability}
\label{s:RealK}

\noindent We shall now present the classical realizability interpretation of the
type system presented in section~\ref{ss:TypingPA2}, following the
method introduced by Krivine~\cite{Kri05}.

First, we shall introduce a calculus of realizers (Krivine's
language~$\lambda_c$) containing the proof-terms of Fig.~\ref{f:PA2},
and give its evaluation rules, that constitute the small-step
operational semantics of the language.
From this, we shall see how to interpret every formula~$A$ of PA2 as a
set of realizers~$|A|$, reading the formula~$A$ as a specification of
the computational behavior of the realizers of~$A$.
The connection between the classical realizability interpretation and
big-step operational semantics in~$\lambda_c$ should become clear in
sections~\ref{s:PrimInt} and~\ref{s:Witness}.

\subsection{A calculus of realizers}
\label{ss:Lamc}

Krivine's language~$\lambda_c$~\cite{Kri05} is a strict extension of
the calculus of proof-terms of \PAtwo\ (section~\ref{ss:TypingPA2}).
The language~$\lambda_c$ actually distinguishes three kinds of
syntactic entities: \emph{terms}, \emph{stacks} and \emph{processes}.
\medbreak\noindent
$\begin{array}{@{}l}
  \textbf{Terms} \\[3pt] \textbf{Stacks} \\[3pt]
  \textbf{Processes} \\
\end{array}$\hfill
$\begin{array}{r@{\quad}r@{\quad}l@{\hskip-10pt}}
  t,u &::=& x \quad|\quad \Lam{x}{t} \quad|\quad tu
  \quad|\quad \kappa \quad|\quad \k_{\pi} \\[3pt]
  \pi &::=& \diamond \quad|\quad t\cdot\pi \\[3pt]
  p,q &::=& t\star \pi
\end{array}$\hfill
$\begin{array}{@{}r}
  (\kappa\in\K) \\[3pt] (t~\text{closed}) \\[3pt] (t~\text{closed}) \\
\end{array}$\medbreak\noindent
Terms of~$\lambda_c$ are pure $\lambda$-terms enriched with two kinds
of constants:
\begin{enumerate}[$\bullet$]
\item\emph{instructions}~$\kappa\in\K$, where~$\K$ is a fixed set of
  constants that contains (at least) an instruction written~$\cc$
  (call$/$cc);
\item\emph{continuation constants} $\k_{\pi}$, one for every
  stack~$\pi$.
\end{enumerate}
Stacks are finite lists of \emph{closed} terms terminated by the stack
constant~$\diamond$.%
\footnote{Krivine allows the formation of stacks using many stack
  constants (representing as many empty stacks), but we will not need
  more than one stack constant here.}
Note that unlike terms (that may be open or closed), stacks only
contain closed terms and are thus closed objects---so that the
continuation constant $\k_{\pi}$ associated to every stack~$\pi$ is
actually a constant.
(The details of the mutual definition of terms and stacks are given
in~\cite{Kri05}.)
Finally, a \emph{process} is simply a pair formed by a closed term~$t$
and a stack~$\pi$.
The set of closed terms (resp.\ the set of stacks) is written
$\Lambda_c$ (resp.\ $\Pi$), and the set of processes is written
$\Lambda_c\star\Pi$.

In section~\ref{s:PrimInt} we shall extend the calculus with extra
instructions to perform fast arithmetic computations.
(See also Remark~\ref{r:DefVsAxiom}.)

\subsubsection{Evaluation}\quad
The set of processes is equipped with a binary relation of one step
\emph{evaluation} written $p\eval p'$, whose reflexive-transitive
closure is written $p\eval^*p'$ as usual.
We assume that this relation satisfies (at least) the following
axioms:
$$\begin{array}{r@{~~}c@{~~}l@{\qquad}l@{\qquad}r@{~~}c@{~~}l}
  \Lam{x}{t}&\star&u\cdot\pi &\eval& t\{x:=u\}&\star&\pi \\
  tu&\star&\pi &\eval& t&\star&u\cdot\pi \\
  \cc&\star&t\cdot\pi &\eval& t&\star&\k_{\pi}\cdot\pi \\
  \k_{\pi}&\star&t\cdot\pi' &\eval& t&\star&\pi \\
\end{array}\leqno\begin{array}{@{}l}
  (\textsc{Grab}) \\ (\textsc{Push}) \\
  (\textsc{Call$/$cc}) \\ (\textsc{Resume}) \\
\end{array}$$
for all $t,u\in\Lambda_c$ and $\pi,\pi'\in\Pi$.
Note that only processes are subject to evaluation: there is no
notion of reduction for either terms or stacks in~$\lambda_c$.

This list of axioms---that basically implements weak head
$\beta$-reduction in presence of the control operator call$/$cc---can
be extended with extra axioms to describe the computational behavior
of the other instructions $\kappa\in\K$.

\begin{rem}\label{r:DefVsAxiom}
  Formally, the definition of the language $\lambda_c$ thus depends on
  two parameters: the set $\K$ of instructions (containing at least
  the instruction~$\cc$), and the relation of evaluation $\succ$
  that fulfils the four axioms given above.
  In particular, the rules (\textsc{Grab}), (\textsc{Push}),
  (\textsc{Call$/$cc}) and (\textsc{Resume}) are only
  \emph{conditions} on the relation $\succ$, but they do not
  constitute a \emph{defi} (by cases) of this relation.
  (The reader is invited to check that these conditions are actually
  the minimal conditions for proving Prop.~\ref{p:KAdequacy}.)
  Putting conditions on the set~$\K$ and on the relation of
  evaluation---rather than defining them completely---naturally makes
  the calculus modular, since this design allows us to enrich the
  calculus with extra instructions (by putting extra conditions
  on~$\K$) and extra evaluation rules (by putting extra conditions
  on~$\succ$), while keeping all the results that have been proved
  using a smaller set of conditions on~$\K$ and~$\succ$%
  \footnote{This is the point of view that is taken
    in~\cite{Kri01,Kri03,Kri05,Kri08}.}.
  Technically, this open design has only one drawback, which is
  that it forbids any form of reasoning by `case analysis' on an
  instruction or on an evaluation step---since the contents of~$\K$
  and the definition of~$\eval$ are not (completely) known.
  Again, the reader is invited to check that this form of reasoning is
  never used in the results presented in Sections~\ref{s:RealK},
  \ref{s:PrimInt} and~\ref{s:Witness}---with the sole exception of
  Lemma~\ref{l:StackExt} in section~\ref{ss:Independence}.
  The set of available instructions and evaluation rules will only be
  closed in section~\ref{s:NegTrans}, in order to define the negative
  translation and to study its properties.
\end{rem}\vfill\eject

\subsection{The realizability interpretation}
\label{ss:KInterp}

\subsubsection{The notion of a pole}\quad
The construction of the classical realizability model is parameterized
by a set of processes $\Bot\subseteq\Lambda_c\star\Pi$, which we
called the \emph{pole} of the model.
We assume that this set is closed under anti-evaluation (or saturated
according to the terminology of~\cite{Kri05}).
Formally:
\begin{defi}
  A \emph{pole} is any set of processes
  $\Bot\subseteq\Lambda_c\star\Pi$ such that the conditions
  $p\eval p'$ and $p'\in\Bot$ together imply $p\in\Bot$ for all
  $p,p'\in\Lambda_c\star\Pi$.
\end{defi}

\begin{rem}
  Since the definition of a pole explicitly depends on the relation of
  evaluation~$\succ$, all the conditions we put on the relation of
  evaluation (see Remark~\ref{r:DefVsAxiom}) are mechanically
  reflected in the definition of the notion of a pole.
  For instance, the rule (\textsc{Push}) is reflected in all
  poles~$\Bot$ by the fact that $t\star u\cdot\pi\in\Bot$ implies
  $tu\star\pi\in\Bot$ (for all terms~$t$, $u$ and for all
  stacks~$\pi$).
  The same holds for the rules (\textsc{Grab}), (\textsc{Call$/$cc})
  and (\textsc{Resume}), as well as for the new rules we shall
  introduce in Section~\ref{s:PrimInt}.
  Putting more conditions on the relation of evaluation
  thus reduces the number of available poles.
\end{rem}

Note that there are two generic ways to define a pole~$\Bot$ from an
arbitrary set of processes $P_0\subseteq\Lambda_c\times\Pi$:
\begin{enumerate}[$\bullet$]
\item The first method is to consider~$P_0$ as a set of
  final (or `accepting') states, and to take~$\Bot$ as the closure
  of~$P_0$ by anti-evaluation, that is: $\Bot=({\eval}P_0)$, which is
  defined by
  $({\eval}P_0)\equiv\{p:\exists p_0\in P_0\,p\eval^*p_0\}$.
\item The second method is to consider~$P_0$ as a set of
  initial (`forbidden') states, and to take~$\Bot$ as the
  \emph{complement} of the closure of~$P_0$ by evaluation, that is:
  $\Bot=(\Lambda_c\star\Pi)\setminus(P_0{\eval})$, where
  $(P_0{\eval})\equiv\{p~:~\exists p_0\in P_0~p_0\eval^*p\}$.
\end{enumerate}
In this paper, we shall build particular poles (in
Section~\ref{s:Witness}) only using the first method, but interesting
uses of the second method can be found in~\cite{Kri03}.

\subsubsection{Truth and falsity values}\quad
From now on, $\Bot$ denotes a fixed pole.
We call a \emph{falsity value} any set of stacks $S\subseteq\Pi$.
By orthogonality, every falsity value $S\subseteq\Pi$ induces a
\emph{truth value} $S^{\Bot}\subseteq\Lambda_c$ defined as:
$$S^{\Bot}~~=~~\{t\in\Lambda_c~:~\forall\pi\,{\in}\,S~~
t\star\pi\in\Bot\}\,.$$

\subsubsection{Valuations and parametric formul{\ae}}\quad
A \emph{valuation} is a function~$\rho$ whose domain is a finite set of
(first- and second-order) variables, such that:
\begin{enumerate}[$\bullet$]
\item $\rho(x)\in\N$ for every first-order variable $x\in\dom(\rho)$;
\item $\rho(X)$ is a (total) function from $\N^k$ to $\P(\Pi)$ (i.e.\
  a \emph{falsity value function}) for every $k$-ary second-order
  variable $X\in\dom(\rho)$.
\end{enumerate}
A \emph{parametric expression} (resp.\ a \emph{parametric formula}) is
simply an arithmetic expression~$e$ (resp.\ a formula~$A$) equipped
with a valuation~$\rho$, that we write $e[\rho]$ (resp.\ $A[\rho]$).
Parametric contexts are defined similarly.
A parametric expression (formula, context) is said to be closed when
every free variable of the underlying expression (formula, context)
belongs to the domain of the attached valuation.

For every closed parametric expression~$e[\rho]$ we write
$\Val(e[\rho])\in\N$ the \emph{value} of $e[\rho]$, interpreting
variables by their images in~$\rho$ while giving to the primitive
recursive function symbols in~$e$ their standard interpretation.

We easily check that:
\begin{lem}\label{l:KConvExprSound}
  If~$e$ and~$e'$ are two arithmetic expressions such that
  $e\conv e'$, then for all valuations~$\rho$ closing~$e$ and~$e'$ we
  have $\Val(e[\rho])=\Val(e'[\rho])$.
\end{lem}

\proof
  By induction on the derivation of~$e\conv e'$.\qed

\subsubsection{The interpretation function}\quad
Every closed parametric formula $A[\rho]$ is interpreted as two sets,
namely: a \emph{falsity value} $\|A[\rho]\|\subseteq\Pi$ and a
\emph{truth value} $|A[\rho]|\subseteq\Lambda_c$.
Both sets are defined by induction on the formula~$A$ as follows:
$$\begin{array}{r@{\quad}c@{\quad}l}
  \|X(e_1,\ldots,e_k)[\rho]\| &=&
  \rho(X)(\Val(e_1[\rho]),\ldots,\Val(e_k[\rho])) \\
  \noalign{\medskip}
  \|\Null(e)[\rho]\| &=& \begin{cases}
    \varnothing &\text{if}~\Val(e[\rho])=0 \\
    \Pi & \text{if}~\Val(e[\rho])\neq 0 \\
  \end{cases} \\
  \noalign{\medskip}
  \|(A\limp B)[\rho]\| &=& |A[\rho]|\cdot\|B[\rho]\| ~~=~~
  \{t\cdot\pi~:~t\in|A[\rho]|,~\pi\in\|B[\rho]\|\} \\
\end{array}$$
%  \noalign{\medskip}
$$\begin{array}{r@{\quad}c@{\quad}l}
  \|(\forall x\,A)[\rho]\| &=&
  \ds\bigcup_{n\in\N}\|A[\rho;x\gets n]\| \\
  \noalign{\medskip}
  \|(\forall X\,A)[\rho]\| &=&
  \ds\bigcup_{\!\!\!\!\!\!F:\N^k\to\P(\Pi)\!\!\!\!\!\!}
  \|A[\rho;x\gets F]\| \\
  \noalign{\medskip}
  |A[\rho]| &=& \|A[\rho]\|^{\Bot}~~=~~
  \{t\in\Lambda_c~:~\forall\pi\in\|A[\rho]\|~~t\star\pi\in\Bot\} \\
\end{array}$$
The reader is invited to check that the sets~$\|A[\rho]\|$
and~$|A[\rho]|$ only depend on the values given by~$\rho$ to the
free variables of~$A$, so that we can drop the valuation~$\rho$
when~$A$ is closed and simply write $\|A\|$ and $|A|$ for
$\|A[\rho]\|$ and $|A[\rho]|$.

We easily check that:
\begin{lem}\label{l:KConvFormSound}
  If~$A$ and~$A'$ are two formul{\ae} of PA2 such that
  $A\conv A'$, then for all valuations~$\rho$ closing~$A$ and~$A'$ we
  have $\|A[\rho]\|=\|A'[\rho]\|$.
\end{lem}

\proof
  By induction on the derivation of~$A\approx A'$ using
  Lemma~\ref{l:KConvExprSound} together with the fact that
  $$\|\bot\|=\bigcup_{S\subseteq\Pi}S=\Pi=
  \|\Null(s(e))[\rho]\|$$
  for all closed parametric expressions $e[\rho]$
  (to interpret $\bot\conv\Null(s(e))$).\qed

Since the truth value~$|A[\rho]|$ and the falsity value~$\|A[\rho]\|$
of the formula~$A$ actually depend on the pole~$\Bot$, we shall
sometimes use the notations $|A[\rho]|_{\Bot}$ and
$\|A[\rho]\|_{\Bot}$ to indicate this dependency explicitly.

\begin{defi}[Realizability]
  Given a pole~$\Bot$, a closed parametric formula~$A[\rho]$ and a
  closed term~$t$, we say that \emph{$t$ realizes~$A[\rho]$} and write
  $t\realK A[\rho]$ when $t\in|A[\rho]|_{\Bot}$, keeping in mind that
  this notation depends on the choice of the particular pole~$\Bot$.
  When $t\in|A[\rho]|_{\Bot}$ for all poles~$\Bot$, we say that
  \emph{$t$ universally realizes~$A[\rho]$} and write
  $t\urealK A[\rho]$.
\end{defi}

\subsubsection{Writing parametric formul{\ae}}\quad
In what follows, we shall often use the convenient shorthand
$$\dot{F}(e_1,\ldots,e_k)~\equiv~(X(e_1,\ldots,e_k))[X\gets F]$$
to denote a parametric formula built from a $k$ary predicate
variable~$X$ that is bound to a particular falsity value
function~$F:\N^k\to\P(\Pi)$ in the attached valuation.
(The dot above the symbol~$\dot{F}$ is here to recall that~$F$ is an
object that belongs to the semantics, not to the syntax.)
By systematically using this notation, we can write parametric
formul{\ae} without explicitly mentioning valuations.
In the sequel, we shall consider (for instance) that the notation
$$\forall z\,(\dot{F}(z)\limp\dot{S})$$
refers to the parametric formula
$$(\forall z\,(X(z)\limp Y))[X\gets F,Y\gets S]$$
where~$X$ and~$Y$ are arbitrarily chosen fresh variables.
Note that the parametric formula defined by such a notation is defined
up to the names of the variables that are bound in the
valuation---but it is easy to see that these names have no impact in
the interpretation of the corresponding parametric formula.

\subsection{The full standard model of PA2 as a degenerate case}
\label{ss:Degenerate}

In the case where~$\Bot=\varnothing$, the classical realizability
model defined above collapses to the full standard model of PA2
(i.e.\ the model where individuals are interpreted by the elements
of~$\N$ and where second-order variables of arity~$k$ are interpreted
by all the subsets of~$\N^k$).
To understand this point, we first notice that when
$\Bot=\varnothing$, the truth value~$S^{\Bot}$ associated to an
arbitrary falsity value $S\subseteq\Pi$ can only take two different
values: $S^{\Bot}=\Lambda_c$ when $S=\varnothing$, and
$S^{\Bot}=\varnothing$ when $S\neq\varnothing$.
Moreover, the realizability interpretation of implication and
universal quantification mimics the standard truth value
interpretation of the corresponding logical construction
(in the case where~$\Bot=\varnothing$).
Writing~$\M$ for the full standard model of PA2, we thus easily show
that:
\begin{lem}
  If $\Bot=\varnothing$, then for every closed formula~$A$ of PA2
  we have
  $$|A|=\begin{cases}
    \Lambda_c &\text{if}~\M\models A \\
    \varnothing &\text{if}~\M\not\models A \\
  \end{cases}$$
\end{lem}

\proof
  We more generally show that for all formul{\ae}~$A$ and for all
  valuations~$\rho$ closing~$A$ (in the sense defined in
  section~\ref{ss:KInterp}) we have
  $$|A[\rho]|=\begin{cases}
    \Lambda_c &\text{if}~\M\models A[\tilde{\rho}] \\
    \varnothing &\text{if}~\M\not\models A[\tilde{\rho}] \\
  \end{cases}$$
  where $\tilde{\rho}$ is the valuation in~$\M$ (in the usual sense)
  defined by
  \begin{enumerate}[$\bullet$]
  \item $\tilde{\rho}(x)=\rho(x)$ if~$x$ is a first-order variable
    such that $x\in\dom(\rho)$;
  \item $\tilde{\rho}(X)=\{(n_1,\ldots,n_k)\in\N^k~:~
    \rho(X)(n_1,\ldots,n_k)=\varnothing\}$ if~$X$ is a second-order
    variable of arity~$k$ such that $X\in\dom(\rho)$.
  \end{enumerate}
  (This characterization is proved by a straightforward induction
  on~$A$.)\qed

An interesting consequence of the above lemma is the following:
\begin{lem}
  If a closed formula~$A$ has a universal realizer $t\urealK A$,
  then~$A$ is true in the full standard model of PA2.
\end{lem}

\proof
  If $t\urealK A$, then $t\in|A|_{\varnothing}$.
  Therefore $|A|_{\varnothing}=\Lambda_c$ and~$\M\models A$.\qed

However, the converse implication is wrong in general, since the
formula $\forall x\,\Nat(x)$ (cf Fig.~\ref{f:PA2}) that expresses the
induction principle over individuals is obviously true in~$\M$, but
has no universal realizer \cite[Theorem~12]{Kri05}%
\footnote{This explains the special treatment of the induction
  principle in section~\ref{ss:Induction}.}.
Nevertheless, the converse implication becomes true when we restrict it
to \emph{arithmetic formul{\ae}}, that is, to the formul{\ae} of the
following language:
$$P,Q~~::=~~ e_1=e_2~~|~~ P\limp Q ~~|~~
\forall x\,(\Nat(x)\limp P)\leqno\textbf{Arithmetic formul{\ae}}$$
(This is a consequence of a slightly more general
result in~\cite[Theorem~21]{Kri05}.)

\begin{rem}
  In the case where~$\Bot\neq\varnothing$, every truth value
  $S^{\Bot}$ is inhabited, for instance by any term of the form
  $\k_{\pi_0}t_0$ where $t_0\star\pi_0\in\Bot$.
  An important consequence of this remark is that a classical realizer
  of a formula~$A$ (w.r.t. to a nonempty pole) can never be taken as a
  `certificate' that the formula~$A$ is true, even when~$A$ is an
  equality.
  (This remark is crucial to understand the specific difficulty of
  witness extraction in classical realizability.)
\end{rem}

\subsection{Adequacy}
\label{ss:KAdequacy}

We call a \emph{substitution} any finite function from proof-variables
to the set $\Lambda_c$ of closed $\lambda_c$-terms, and we denote by
$t[\sigma]$ the term obtained by applying a substitution~$\sigma$ to a
term~$t$.
Given a substitution $\sigma$ and a closed parametric context
$\Gamma[\rho]$, we write $\sigma\realK\Gamma[\rho]$ when the following
conditions are fulfilled:
\begin{enumerate}[(1)]
\item $\dom(\Gamma)\subseteq\dom(\sigma)$;
\item $\sigma(x)\realK A[\rho]$ for every declaration
  $(x:A)\in\Gamma$.
\end{enumerate}
We say that:
\begin{enumerate}[$\bullet$]
\item A judgment $\TYPK{\Gamma}{t}{A}$ is \emph{sound} (w.r.t.\ the
  pole~$\Bot$) when for all valuations~$\rho$ and for all
  substitutions~$\sigma$ such that $\sigma\realK\Gamma[\rho]$, we have
  $t[\sigma]\realK A[\rho]$.
\item An inference rule $\frac{P_1\cdots P_n}{C}$
  (where~$P_1,\ldots,P_n$ and $C$ are typing judgments) is
  \emph{sound} (w.r.t.\ the pole~$\Bot$) when the soundness of its
  premises $P_1,\ldots,P_n$ (in the above sense) implies the soundness
  of its conclusion~$C$.
\end{enumerate}
From these definitions, it is clear that the conclusion of any
typing derivation formed with only sound inference rules is sound
w.r.t.\ all poles~$\Bot$.

\begin{prop}[Adequacy]\label{p:KAdequacy}
  The typing rules of \PAtwo\ (Fig.~\ref{f:PA2}) are sound w.r.t.\
  all poles $\Bot\subseteq\Lambda_c\times\Pi$.
\end{prop}

\proof
  The soundness of the introduction rule of~$\top$ is obvious
  (since $|\top|=\Lambda_c$) and the soundness of the conversion rule
  follows from Lemma~\ref{l:KConvFormSound}.
  The soundness of the remaining typing rules is proved
  in~\cite{Kri05}.\qed

A consequence of this proposition is that closed proof-terms that are
built using the type system of \PAtwo\ are actually universal
realizers of the corresponding formul{\ae}.
(But not all realizers can be detected via typing~\cite{Kri05}.)

\section{Primitive natural numbers}
\label{s:PrimInt}

\noindent Through the formul{\ae}-as-types paradigm, the relativized form of
first-order universal quantification $\forall x\,(\Nat(x)\limp A(x))$
corresponds to the (dependent) type of all functions mapping realizers
of the formula $\Nat(n)$ to realizers of the formula $A(n)$ for every
$n\in\N$.
To get a realizer of the formula $A(n)$ (for a particular value of
$n\in\N$) from a realizer~$t$ of the formula
$\forall x\,(\Nat(x)\limp A(x))$, 
it suffices to apply the term~$t$ to the Church numeral
$\Lam{xf}{f^nx}$ using the following fact
\begin{fact}
  For every $n\in\N$ one has:
  $\TYPK{}{\Lam{xf}{f^nx}}{\Nat(n)}$.
\end{fact}
\noindent combined with the property of adequacy (Prop.~\ref{p:KAdequacy}).

On the other hand, Church numeral $\Lam{xf}{f^nx}$ is far from being
the only realizer of the formula $\Nat(n)$---the situation being
much more complex than in intuitionistic realizability due to the
presence of continuations in realizers.
However, it is always possible to effectively retrieve (in some sense)
the natural number~$n$ from an arbitrary realizer of the 
formula $\Nat(n)$, and the traditional way to achieve this in
classical realizability is to use a storage
operator~\cite{Kri94,Kri05}.
We propose here another method by changing the representation of
numerals.

Indeed, the main defect of Church numerals is not only their very poor
efficiency in practical computations (especially for large values),
but also the non atomicity of their encoding that makes them very hard
to track through a negative translation towards intuitionistic logic.
%such as the one we propose to define in section~???.
For this reason, we present here an alternative implementation of
natural numbers in classical realizability, based on the introduction
of specific constants to represent natural numbers with new
instructions to compute with them.

\subsection{Extending the language of realizers}
\label{ss:KRealizExt}

We now enrich%
\footnote{See remark~\ref{r:DefVsAxiom} p.~\pageref{r:DefVsAxiom}.}
the instruction set~$\K$ with the following constants:
\begin{enumerate}[$\bullet$]
\item For every $n\in\N$, a constant $\widehat{n}\in\K$ representing
  the natural number~$n$ as a pure datum.
  Here, the constant $\widehat{n}$ hardly deserves the name of an
  instruction, since it comes with no evaluation rule.
  The intuition is that the constant~$\widehat{n}$ is only meaningful
  as a datum in the stack, not in head position.%
  \footnote{This is similar to the situation in most programming
    languages, where numbers are represented using machine numbers (or
    blocks of machine numbers) that are meaningless as pointers, so
    that executing them usually raises a memory fault.}
\item Two constants~$\Succ$ and~$\Rec$ with the evaluation rules
  $$\begin{array}{r@{~~}c@{~~}l@{\qquad}l@{\qquad}r@{~~}c@{~~}l}
    \Succ&\star&\widehat{n}\cdot u\cdot\pi &\eval&
    u&\star&\widehat{n+1}\cdot\pi \\
    \Rec&\star&u_0\cdot u_1\cdot\widehat{0}\cdot\pi &\eval&
    u_0&\star&\pi \\
    \Rec&\star&u_0\cdot u_1\cdot\widehat{n+1}\cdot\pi &\eval&
    u_1&\star&\widehat{n}\cdot
    (\Rec\,u_0\,u_1\,\widehat{n})\cdot\pi\\
  \end{array}\leqno\begin{array}{@{}l}
    (\textsc{Succ})\\(\textsc{Rec-0})\\(\textsc{Rec-S})\\
  \end{array}$$
  for all $u,u_0,u_1\in\Lambda_c$, $n\in\N$ and $\pi\in\Pi$.
\end{enumerate}
With these new instructions, it is more generally possible to
implement every recursive function $f$ of arity~$k$ as a term
$\check{f}$ with the reduction rule
$$\check{f}\star\widehat{n}_1\cdots\widehat{n}_k\cdot u\cdot\pi
\quad\eval^*\quad u\star\widehat{m}\cdot\pi\,,$$
for all $(n_1,\ldots,n_k)\in\dom(f)$, writing~$m=f(n_1,\ldots,n_k)$.%
\footnote{In the case we want to go beyond primitive recursion, it is
  necessary to use a fixpoint combinator to implement minimization.}
To improve efficiency, we can also introduce the $\check{f}$s (or some
of them) as new instructions.

Apart from the representation of numerals as pure data, every natural
number $n\in\N$ can be also represented as a
\emph{program}~$\check{n}$ defined by
$\check{n}\equiv\Lam{x}{x\widehat{n}}$.
(We will momentarily see how to give a `type' to this program in PA2.)

\subsection{Extending the realizability interpretation}
\label{ss:ExtRealK}

To understand the computational behavior of the instructions that come
with our alternative representation of numerals, we extend the
language of formul{\ae} of~\PAtwo\ with a new syntactic construct
$\{e\}\limp B$ where~$e$ is an arithmetic expression and~$B$ a
formula.
(This extension is part of a larger system~$\PAext$ that will be
introduced in section~\ref{ss:PA2ext}.)
Intuitively, this formula corresponds to the type of all functions
taking the representation of the value of~$e$ as the
constant~$\widehat{n}$ (where $n=\Val(e)$) and return an object of
type~$B$.

Formally, the realizability interpretation of the formul{\ae}
of~$\PAtwo$ (section~\ref{ss:KInterp}) is extended to the syntactic
construct $\{e\}\limp B$ by letting:
$$\begin{array}{rcl}
  \|(\{e\}\limp B)[\rho]\| &=&
  \{\widehat{n}\cdot\pi~:~n=\Val(e[\rho]),~\pi\in\|B[\rho]\|\}\,. \\
\end{array}$$
In this extended syntax, we can now give a type to the lazy numeral
$\check{n}\equiv\Lam{x}{x\widehat{n}}$ by letting
$\Nat'(e)\equiv\forall Z\,((\{e\}\limp Z)\limp Z)$ and checking that:
\begin{lem}
  For every $n\in\N$:\quad
  $\check{n}~\equiv~\Lam{x}{x\widehat{n}}~\urealK~\Nat'(n)$
\end{lem}

\proof
  Let~$\Bot$ be a fixed pole, and consider an arbitrary element
  of falsity value
  $\|\Nat'(n)\|=\|\forall Z\,((\{n\}\limp Z)\limp Z)\|$, that
  is: a stack of the form $u\cdot\pi$ where
  $u\in|\{n\}\limp\dot{S}|$ and $\pi\in S$ for some falsity
  value $S\in\P(\Pi)$.
  We have $\Lam{x}{x\widehat{n}}\star u\cdot\pi\eval^*
  u\star\widehat{n}\cdot\pi$.
  But since $\widehat{n}\cdot\pi\in\|\{n\}\limp\dot{S}\|$ we get
  $u\star\widehat{n}\cdot\pi\in\Bot$, hence
  $\Lam{x}{x\widehat{n}}\star u\cdot\pi\in\Bot$ by
  anti-evaluation.\qed

Moreover:
\begin{lem}\label{l:TypingSuccRec}
  Writing $\forallN x\,A(x)\equiv\forall x\,(\{x\}\limp A(x))$,
  we have:
  \begin{enumerate}[\em(1)]
  \item $\Succ~\urealK~\forallN x\,\Nat'(s(x))$
  \item $\Rec~\urealK~\forall Z\,\bigl(Z(0)\limp
    \forallN y\,(Z(y)\limp Z(s(y)))\limp\forallN x\,Z(x)\bigr)$
  \end{enumerate}
\end{lem}

\proof
  Let~$\Bot$ be a fixed pole.
  \begin{enumerate}[(1)]
  \item Let us consider an arbitrary element of
    $\|\forallN x\,\Nat'(s(x))\|$, that is: a stack of the form
    $\widehat{n}\cdot u\cdot\pi$, where $n\in\N$,
    $u\in|\{s(n)\}\limp\dot{S}|$ and $\pi\in S$ for some falsity value
    $S\subseteq\Pi$.
    We want to show that
    $\Succ\star\widehat{n}\cdot u\cdot\pi\in\Bot$.
    Using the evaluation rule of~$\Succ$, we get
    $\Succ\star\widehat{n}\cdot u\cdot\pi\eval
    u\star\widehat{n+1}\cdot\pi$.
    But since $\widehat{n+1}\cdot\pi\in\|\{s(n)\}\limp\dot{S}\|$, we
    have $u\star\widehat{n+1}\cdot\pi\in\Bot$, hence
    $\Succ\star\widehat{n}\cdot u\cdot\pi\in\Bot$ by anti-evaluation.
  \item Let us take a falsity value function $F:\N\to\P(\Pi)$ and
    consider two realizers $u_0\in|\dot{F}(0)|$ and
    $u_1\in|\forallN y\,(\dot{F}(y)\limp\dot{F}(s(y)))|$.
    We first show by induction on~$n\in\N$ that for all stacks
    $\pi\in F(n)$ we have
    $\Rec\star u_0\cdot u_1\cdot\widehat{n}\cdot\pi\in\Bot$.
    \begin{enumerate}[$-$]
    \item Base case.\quad Take $\pi\in F(0)$.
      We have $\Rec\star u_0\cdot u_1\cdot\widehat{0}\cdot\pi
      \eval u_0\star\pi\in\Bot$ (since $u_0\in F(0)^{\Bot}$),
      hence $\Rec\star u_0\cdot u_1\cdot\widehat{0}\cdot\pi\in\Bot$
      by anti-evaluation.
    \item Let us assume that the property holds for $n\in\N$,
      and consider a stack $\pi\in F(n+1)$.
      We have $\Rec\star u_0\cdot u_1\cdot\widehat{n+1}\cdot\pi
      \eval u_1\star\widehat{n}\cdot
      (\Rec\,u_0\,u_1\,\widehat{n})\cdot\pi$.
      We now want to show that
      $\Rec\,u_0\,u_1\,\widehat{n}\in|\dot{F(n)}|$.
      For that, we take a stack $\pi'\in F(n)$ and get
      $\Rec\,u_0\,u_1\,\widehat{n}\star\pi'\eval^*
      \Rec\star u_0\cdot u_1\cdot\widehat{n}\cdot\pi'\in\Bot$
      by induction hypothesis, hence
      $\Rec\,u_0\,u_1\,\widehat{n}\star\pi'\in\Bot$ by
      anti-evaluation.
      Thus we have
      $\Rec\,u_0\,u_1\,\widehat{n}\in|\dot{F(n)}|$,
      hence we get
      $$\begin{array}{rcl}
        \widehat{n}\cdot(\Rec\,u_0\,u_1\,\widehat{n})\cdot\pi
        &~\in~&\|\{n\}\limp\dot{F}(n)\limp\dot{F}(s(n))\|\\
        &&{\subseteq}~
        \|\forallN y\,(\dot{F}(y)\limp\dot{F}(s(y)))\|\,.
      \end{array}$$
      Therefore $u_1\star\widehat{n}\cdot
      (\Rec\,u_0\,u_1\,\widehat{n})\cdot\pi\in\Bot$, and
      $\Rec\star u_0\cdot u_1\cdot\widehat{n+1}\cdot\pi\in\Bot$
      by anti-evaluation.
    \end{enumerate}
    We have shown that
    $\Rec\star u_0\cdot u_1\cdot\widehat{n}\cdot\pi\in\Bot$
    for all $F:\N\to\P(\Pi)$ and for all $u_0\in|\dot{F}(0)|$,
    $u_1\in|\forallN y\,(\dot{F}(y)\limp\dot{F}(s(y)))|$,
    $n\in\N$ and $\pi\in F(n)$.
    But this precisely means that $\Rec$ realizes the desired
    formula.\qed
  \end{enumerate}

\subsection{Extending the type system}
\label{ss:PA2ext}

To facilitate the construction of universal realizers using the new
instructions, we define an extension of~$\PAtwo$ (\ref{f:PA2}), which
we call~$\PAext$.
The specific formation rules and typing rules of this system are
summarized in Fig.~\ref{f:PA2ext}.

\begin{figure}[ht!]
  \def\myskip{\vskip 9pt}
  $$\begin{array}{@{}c@{}}
    \hline\hline
    \noalign{\medskip}
    \underline{\textbf{Syntactic constructs}} \\
    \noalign{\bigskip}
    \begin{array}{l@{\qquad}r@{\quad}r@{\quad}l}
      \textbf{Formul{\ae}}
      & A,B &::=& \cdots \quad|\quad \{e\}\limp B \\[3pt]
      \textbf{Proof-terms}
      & t,u &::=& \cdots \quad|\quad \widehat{n} \quad|\quad
      \Succ \quad|\quad \Rec \\[3pt]
      \textbf{Contexts}
      & \Gamma &::=& \cdots \quad|\quad \Gamma,x:\{e\} \\
    \end{array} \\
    \noalign{\bigskip}
    \underline{\textbf{Abbreviations}} \\
    \noalign{\bigskip}
    \begin{array}{l@{}r@{\quad}r@{\quad}l}
      & \Nat'(e) &\equiv& \forall Z\,((\{e\}\limp Z)\limp Z) \\
      & \forallN x\,A &\equiv& \forall x\,(\{x\}\limp A) \\
      & \existsN x\,A &\equiv& \forall Z\,
      (\forall x\,(\{x\}\limp A\limp Z)\limp Z) \\
    \end{array} \\
    \noalign{\bigskip}
    \underline{\textbf{Typing rules}} \\
    \noalign{\bigskip}
    \infer{\TYPK{\Gamma}{\Rec}
      {\forall Z\,(Z(0)\limp\forallN y\,(Z(y)\limp Z(s(y)))
        \limp\forallN x\,Z(x))}}{} \\
    \noalign{\myskip}
    \infer{\TYPK{\Gamma}{\Succ}
      {\forallN x~\Nat'(s(x))}}{} \qquad\qquad
    \infer{\TYPK{\Gamma}{\Lam{x}{t}}{\{e\}\limp B}}{
      \TYPK{\Gamma,x:\{e\}}{t}{B}
    } \\
    \noalign{\myskip}
    \infer[\scriptstyle(x:\{e\})\in\Gamma]
    {\TYPK{\Gamma}{t\,x}{B}}{
      \TYPK{\Gamma}{t}{\{e\}\limp B}
    } \quad~~
    \infer{\TYPK{\Gamma}{t\,\widehat{n}}{B}}{
      \TYPK{\Gamma}{t}{\{n\}\limp B}
    } \\
    \noalign{\medskip}
    \hline\hline
  \end{array}$$\vspace{-12pt}
  \caption{Extending PA2 with primitive numerals}
  \label{f:PA2ext}
\end{figure}

\noindent Compared to~$\PAtwo$, the grammar of the formul{\ae} of~$\PAext$ is
enriched with the syntactic construct~$\{e\}\limp B$ introduced in
Section~\ref{ss:ExtRealK}.
(Arithmetic expressions remain unchanged.)
To reflect the presence of a second form of implication, typing
contexts of system~$\PAext$ introduce a second form of declaration,
written $x:\{e\}$, that expresses that the proof-variable~$x$ is bound
to the constant~$\widehat{n}$, where~$n$ is the value of~$e$ (i.e.\
$n=\Val(e)$).

Proof-terms of~$\PAext$ are the proof-terms of~$\PAtwo$ enriched
with the constants~$\widehat{n}$ (for all $n\in\N$), $\Succ$
and~$\Rec$.
System~$\PAext$ provides typing rules for the constants~$\Succ$
and $\Rec$, as well as an introduction rule and two elimination rules
for the formula $\{e\}\limp B$. 
Note that in this system, we can only apply a proof-term of type
$\{e\}\limp B$ to a variable (declared with $x:\{e\}$) or to a
constant of the form~$\widehat{n}$---in which case we must have
$e\equiv s^n(0)$.

\subsubsection{The realizability interpretation of~$\PAext$}\quad
The realizability interpretation of~$\PAext$ is defined as
for~$\PAtwo$, using the interpretation of the formula~$\{e\}\limp B$
described in Section~\ref{ss:ExtRealK}.
(Of course, we now work with a set~$\K$ of instructions and a relation
of evaluation that fulfill the conditions given in
Section~\ref{s:RealK} and~\ref{s:PrimInt}.)

To express the soundness of the new typing rules, we first have to
adapt the definition of $\sigma\realK\Gamma[\rho]$ to the extended
notion of context.
For that, we say that a substitution~$\sigma$ realizes a closed
parametric context $\Gamma[\rho]$ and write $\sigma\realK\Gamma[\rho]$
when the following conditions are fulfilled:
\begin{enumerate}[(1)]
\item $\dom(\Gamma)\subseteq\dom(\sigma)$;
\item $\sigma(x)\realK A[\rho]$ for every declaration
  $(x:A)\in\Gamma$;
\item $\sigma(x)=\widehat{n}$ where $n=\Val(e[\rho])$ for every
  declaration $(x:\{e\})\in\Gamma$.
\end{enumerate}
(This definition obviously coincides with the former definition in the
case where the context~$\Gamma$ only contains declarations of the form
$(x:A)$.)
The definition of sound judgments and of sound valid rules (w.r.t.\ a
fixed pole) immediately extends to the new system, so that we can
check the following:
\begin{prop}[Adequacy]\label{p:KAdequacyExt}
  The typing rules of \PAext\ are sound w.r.t.\ all poles
  $\Bot\subseteq\Lambda_c\times\Pi$.
\end{prop}

\proof
  Let~$\Bot$ be a pole.
  We only treat the specific rules of~$\PAext$ (Fig.~\ref{f:PA2ext}).
  \begin{enumerate}[$\bullet$]
  \item Typing rules for~$\Succ$ and~$\Rec$:
    immediately follows from Lemma~\ref{l:TypingSuccRec}.
  \item Introduction rule of $\{e\}\limp B$.
    Let us assume that $\TYPK{\Gamma,x:\{e\}}{t}{B}$ is sound.
    To show that the judgment
    $\TYPK{\Gamma}{\Lam{x}{t}}{\{e\}\limp B}$
    is sound too, consider a valuation~$\rho$ with a
    substitution~$\sigma$ such that $\sigma\realK\Gamma[\rho]$.
    We want to prove that
    $(\Lam{x}{t})[\sigma]\in|(\{e\}\limp B)[\rho]|$.
    For that, let us consider an arbitrary element of
    $\|(\{e\}\limp B)[\rho]\|$,
    that is: a stack of the form $\widehat{n}\cdot\pi$ where
    $n=\Val(e[\rho])$ and $\pi\in\|B[\rho]\|$, and let us prove
    that $(\Lam{x}{t})[\sigma]\star n\cdot\pi\in\Bot$.
    Let $\sigma'=(\sigma,x:=\widehat{n})$.
    We have $\sigma'\Vdash(\Gamma,x:\{e\})[\rho]$, hence
    $t[\sigma']\in|B[\rho]|$ from the soundness of judgment
    $\TYPK{\Gamma,x:\{e\}}{t}{B}$.
    By evaluating the process $(\Lam{x}{t})[\sigma]\star n\cdot\pi$ we
    get $(\Lam{x}{t})[\sigma]\star n\cdot\pi\eval
    t[\sigma']\star\pi\in\Bot$ (since $t[\sigma']\in|B[\rho]|$), hence
    $(\Lam{x}{t})[\sigma]\star n\cdot\pi\in\Bot$ by anti-evaluation.
  \item Elimination rules of $\{e\}\limp B$:
    both cases are straightforward.\qed
  \end{enumerate}

\noindent Thanks to this extension, it is easy to check (by means of typing)
that the new relativization predicate $\Nat'(x)$ is logically
equivalent (in~$\PAext$) to the traditional relativization predicate
$\Nat(x)$ defined in section~\ref{ss:Induction}:
$$\begin{array}{r@{\quad}c@{\quad}l}
  \Lam{z}{z\,\check{0}\,(\Lam{y}{y\,\check{s}})}&:&
  \forall x\,(\Nat(x)\limp\Nat'(x)) \\
  \Lam{z}{z\,(\Rec\,(\Lam{xf}{x})\,(\Lam{\_nxf}{f\,(n\,x\,f)}))}
  &:& \forall x\,(\Nat'(x)\limp\Nat(x)) \\
\end{array}$$
(Intuitively, the above terms convert a Church numeral into the
corresponding lazy numeral and vice-versa.)
Moreover, we can check that the formula $\forallN x\,A(x)$ defined by
the shorthand
$$\forallN x\,A(x)~\equiv~\forall x\,(\{x\}\limp A(x))
\eqno(\text{nat-as-data relativization})$$
is logically equivalent to the formula
$$\forall x\,(\Nat'(x)\limp A(x))
\eqno(\text{nat-as-program relativization})$$
by means of the following proof-terms:
$$\begin{array}{r@{\quad}c@{\quad}l}
  \Lam{fx}{f\,(\Lam{y}{yx})}
  &:& \forall x\,(\Nat'(x)\limp A(x))~\limp~\forallN x\,A(x) \\
  \Lam{fx}{xf}
  &:& \forallN x\,A(x)~\limp~\forall x\,(\Nat'(x)\limp A(x)) \\
\end{array}$$
(Intuitively, functions of type $\forallN x\,A(x)$ expect a fully
computed natural number represented as a datum on the top of the
stack, whereas functions of type $\forall x\,(\Nat'(x)\limp A(x))$
expect a lazy representation of a natural number on the top of the
stack, whose corresponding value can be computed later.)

The same remark holds for the two different ways to relativize
first-order existential quantification using primitive numerals
$$\begin{array}{l}
  \forall Z\,(\forall x\,(\Nat'(x)\limp A(x)\limp Z)\limp Z)\\[3pt]
  \existsN x\,A(x)\equiv
  \forall Z\,(\forall x\,(\{x\}\limp A(x)\limp Z)\limp Z)\\[3pt]
\end{array}$$
that are provably equivalent.

In what follows, we shall thus only consider the problem of witness
extraction from universal realizers of existential formul{\ae} of the
form $\existsN x\,A(x)$, whose witnesses are the most directly
accessible.

\section{Witness extraction in classical realizability}
\label{s:Witness}

\noindent In this section, we are interested in the problem of extracting a
witness of a closed existential formula $\existsN x\,A(x)$ from a
fixed universal realizer~$t_0$ of this formula:
$$t_0~\urealK~\existsN x\,A(x)~\equiv~
\forall Z\,(\forall x\,(\{x\}\limp A(x)\limp Z)\limp Z)\,.$$
(As a particular case, $t_0$ may be a proof term of $\existsN x\,A(x)$
in PA2.)

Throughout this section, we assume that the instruction set~$\K$
contains (at least) the extra instructions~$\widehat{n}$, $\Succ$ and
$\Rec$ presented in Section~\ref{ss:KRealizExt}, with their
accompanying rules.
For convenience, we also assume the existence of an instruction
$\Stop$ with no evaluation rule, that is intended to abort computation
once the desired witness has been found.
However, the proofs of Prop.~\ref{p:ExtrNaive}, \ref{p:ExtrSigma01}
and~\ref{p:ExtrDec} do not rely on any particular assumption
on~$\Stop$, so that these propositions still hold if we consider
that~$\Stop$ denotes a fixed closed $\lambda_c$-term.

The witness extraction methods discussed in
Sections~\ref{ss:ExtrSigma01} and~\ref{ss:ExtrDec} are directly
inspired from the techniques presented in~\cite{Kri05}, while the
method presented in Section~\ref{ss:ExtrKam} is due to the author.

\subsection{The failure of the naive method}
\label{ss:ExtrNaive}

To extract a witness from the universal realizer
$t_0\urealK\existsN x\,A(x)$,
a natural idea would be to apply~$t_0$ to the term
$\Lam{xy}{\Stop\,x}$ that extracts the first component of the
`pair'~$t_0$ and passes it to~$\Stop$.
Applying this idea, we get the following:
\begin{prop}\label{p:ExtrNaive}
  For all~$\pi\in\Pi$, the process
  $t_0\star(\Lam{xy}{\Stop\,x})\cdot\pi$
  evaluates (in a finite number of steps) to a process of the form
  $\Stop\star\widehat{n}\cdot\pi$ for some $n\in\N$.
\end{prop}

\proof
  Let us take a stack~$\pi\in\Pi$ and work in the pole defined by
  $$\Bot~=~\{p~:~\exists n\,{\in}\,\N~~
  p\eval^*\Stop\star\widehat{n}\cdot\pi\}\,.$$
  Writing $S=\{\pi\}$, we easily check that
  $\Stop\realK\forall x\,(\{x\}\limp\dot{S})$ (from the definition
  of~$\Bot$), so that
  $\Lam{xy}{\Stop\,x}\realK\forall x\,(\{x\}\limp A(x)\limp\dot{S})$
  (by Prop.~\ref{p:KAdequacy}).
  Therefore $(\Lam{xy}{\Stop\,x})\cdot\pi\in\|\existsN x\,A(x)\|$,
  and thus $t_0\star(\Lam{xy}{\Stop\,x})\cdot\pi\in\Bot$.\qed

Alas, this result gives us no warranty that the natural number~$n$ we
get by this method is such that $A(n)$ is true (in the full standard
model).
The mistake here is that we have dropped the second component~$y$ of
the pair~$t_0$ (that cannot be taken as a certificate that $A(n)$
holds), and we shall momentarily see that this component is actually
the crucial ingredient of the extraction process.

\subsection{Extraction in the $\Sigma^0_1$-case}
\label{ss:ExtrSigma01}

Let us now consider the particular case where the predicate $A(x)$ is
of the form $A(x)\equiv f(x)=0$, where~$f$ is a unary function symbol
of the signature corresponding to (and denoted by) a primitive
recursive function still written~$f$.

To understand how to extract a (correct) witness from~$t_0$ in this
case, let us first study the denotation of equalities in the
realizability model:
\begin{lem}\label{l:RealizEq}
  Let~$e_1$ and~$e_2$ be closed arithmetic expressions.
  For all poles~$\Bot$ we have
  $$\|e_1=e_2\|~=~\begin{cases}
    \{t\cdot\pi~:~(t\star\pi)\in\Bot\}~=~\|\mathbf{1}\| &
    \text{if}~~\Val(e_1)=\Val(e_2) \\
    \Lambda\cdot\Pi~=~\|\top\limp\bot\| &
    \text{if}~~\Val(e_1)\neq\Val(e_2) \\
  \end{cases}$$
  (writing $\mathbf{1}\equiv\forall Z\,(Z\limp Z)$).
\end{lem}

In other words, true equalities are interpreted the same way as the
formula $\mathbf{1}\equiv\forall Z\,(Z\limp Z)$ whereas false
equalities are interpreted the same way as the formula $\top\limp\bot$
in the classical realizability model.
If $u$ is a realizer of the formula $f(n)=0$ (w.r.t.\ a particular
pole~$\Bot$), then we can distinguish two cases:
\begin{enumerate}[$\bullet$]
\item The equality $f(n)=0$ is true.
  In this case, we can think of~$u$ (${\realK}~\mathbf{1}$) as a term
  that essentially behaves as the identity term 
  $\Lam{z}{z}$: when coming in head position, it simply vanishes
  and gives the control to its argument.
\item The equality $f(n)=0$ is false.
  In this case, we can think of~$u$ (${\realK}~\top\limp\bot$) as a
  term that consumes its argument (whatever it is) and then backtracks
  to an earlier point in the computation.%
\end{enumerate}
Of course, this informal description is only an loose approximation of
the actual behavior of the realizer $u\realK f(n)=0$ (which may
considerably vary depending on the choice of~$\Bot$), but it gives us
the clue to fix the naive extraction method.

The idea is to apply the universal realizer
$t_0\urealK\existsN x\,f(x)=0$ to the term $\Lam{xy}{y\,(\Stop\,x)}$
that inserts a `breakpoint'~$y$ before returning $x$.
If the first component~$x$ is a correct witness, then the second
component~$y$ will vanish and let the program return the correct
answer.
If the first component~$x$ is incorrect, then~$y$ will issue a
backtrack, and this until a correct witness has been found.

We can now formalize this intuition as follows:
\begin{prop}\label{p:ExtrSigma01}
  For all $\pi\in\Pi$, the process
  $t_0\star(\Lam{xy}{y\,(\Stop\,x)})\cdot\pi$
  evaluates (in a finite number of steps) to a process of the form
  $\Stop\star\widehat{n}\cdot\pi$ for some natural number $n\in\N$
  such that $f(n)=0$.
\end{prop}

\proof
  Let us take a stack~$\pi\in\Pi$ and work in the pole defined by
  $$\Bot~=~\{p~:~\exists n\,{\in}\,\N~~(f(n)=0~~\text{and}~~
  p\eval^*\Stop\star\widehat{n}\cdot\pi)\}\,.$$
  Writing $S=\{\pi\}$, we easily check that
  $\Stop\realK\{n\}\limp\dot{S}$ for all $n\in\N$ such that $f(n)=0$
  (from the very definition of~$\Bot$ and~$S$).
  Let us now show that the term $\Lam{xy}{y\,(\Stop\,x)}$ realizes the
  formula $\forall x\,(\{x\}\limp f(x)=0\limp\dot{S})$.
  For that, consider an arbitrary element of the falsity value of this
  formula, that is: a stack of the form $\widehat{n}\cdot u\cdot\pi$
  for some $n\in\N$ and $u\in|f(n)=0|$.
  We have
  $$\Lam{xy}{y\,(\Stop\,x)}\star\widehat{n}\cdot u\cdot\pi
  ~\eval^*~u\star(\Stop\,\widehat{n})\cdot\pi\,.$$
  To show that $u\star(\Stop\,\widehat{n})\cdot\pi\in\Bot$,
  we distinguish two cases:
  \begin{enumerate}[$\bullet$]
  \item $f(n)=0$.\
    In this case, we have $\Stop\,\widehat{n}\realK\dot{S}$ (using the
    `type' we gave to $\Stop$), hence
    $(\Stop\,\widehat{n})\star\pi\in\Bot$ and thus
    $(\Stop\,\widehat{n})\cdot\pi\in\|\mathbf{1}\|=\|f(n)=0\|$
    (by Lemma~\ref{l:RealizEq}).
    Therefore $u\star(\Stop\,\widehat{n})\cdot\pi\in\Bot$.
  \item $f(n)\neq 0$.\
    In this case, we have
    $(\Stop\,\widehat{n})\cdot\pi\in\|\top\limp\bot\|=\|f(n)=0\|$,
    hence $u\star(\Stop\,\widehat{n})\cdot\pi\in\Bot$.
  \end{enumerate}
  In both cases we deduce that
  $\Lam{xy}{y\,(\Stop\,x)}\star\widehat{n}\cdot u\cdot\pi\in\Bot$ by
  anti-evaluation, which finishes the proof that
  $\Lam{xy}{y\,(\Stop\,x)}\realK
  \forall x\,(\{x\}\limp f(x)=0\limp\dot{S})$.
  From the latter we deduce that
  $(\Lam{xy}{y\,(\Stop\,x)})\cdot\pi\in\|\existsN x\,f(x)=0\|$,
  so that
  $t_0\star(\Lam{xy}{y\,(\Stop\,x)})\cdot\pi\in\Bot$.\qed

\begin{rem}\label{r:Print}
  The simple (and reliable) extraction procedure presented above
  returns a correct witness without keeping track of the intermediate
  witnesses proposed by the realizer~$t_0$.
  A simple way to display them during the computation is to introduce
  an instruction $\Print$ such that
  $$\Print\star\widehat{n}\cdot u\cdot\pi~\eval~u\star\pi
  \eqno(n\in\N,~u\in\Lambda,~\pi\in\Pi)$$
  while printing the natural number~$n$ on some output device
  (the second part of the specification of $\Print$ being purely
  informal).
  From the only evaluation rule of~$\Print$, we easily check that
  $\Print\urealK\forall x\,(\{x\}\limp\mathbf{1})$.
  It is then a straightforward exercise to adapt the proof of
  Prop.~\ref{p:ExtrSigma01} when the process
  $t_0\star(\Lam{xy}{y\,(\Stop\,x)})\cdot\pi$ is replaced by the
  process $t_0\star(\Lam{xy}{\Print\,x\,y\,(\Stop\,x)})\cdot\pi$
  that ultimately does the same job---while printing the intermediate
  results.
\end{rem}

In section~\ref{ss:Compare} we shall reinterpret the witness
extraction method of Prop.~\ref{p:ExtrSigma01} through a well-suited
negative translation.

\subsection{Independence of the witness w.r.t. the stack~$\pi$}
\label{ss:Independence}

It is easy to see that the witness computed by the process
$t_0\star(\Lam{xy}{y\,(\Stop\,x)})\cdot\pi$ (in the sense of
Prop.~\ref{p:ExtrSigma01}) does not actually depend on the
stack~$\pi$, provided we make the following `closed world'
assumptions:
\begin{enumerate}[(1)]
\item The relation of (one step) evaluation~$\succ$ is defined
  as the union of the rules (\textsc{Grab}), (\textsc{Push}),
  (\textsc{Call$/$cc}), (\textsc{Resume}), (\textsc{Succ}),
  (\textsc{Rec-0}) and (\textsc{Rec-S}) (cf
  Sections~\ref{ss:Lamc} and~\ref{ss:KRealizExt}).
  In particular, evaluation is deterministic.
  \smallbreak
\item The term~$t_0$ contains no continuation
  constant~$\k_{\pi}$, that is: $t_0$ is a \emph{proof-like term}
  according to the terminology of~\cite{Kri05}.
  Note that this condition is automatically fulfilled when~$t_0$
  is a proof-term built in system~\PAext.
  \smallbreak
\item The term~$\Stop$ is an extra instruction (with no
  evaluation rule).
\end{enumerate}

To prove the desired independence result, we define an operation of
\emph{stack extension} for terms, stacks and processes as follows%
\footnote{For an account of the possible uses of this technique,
  see~\cite{GuiPhD}.}.
Given a fixed stack~$\pi_0$, we denote by
$t\{\diamond:=\pi_0\}$ (resp.\ $\pi\{\diamond:=\pi_0\}$,
$p\{\diamond:=\pi_0\}$) the term~$t$ (resp.\ the stack~$\pi$, the
process~$p$) in which every occurrence of the stack bottom~$\diamond$
is replaced by the stack~$\pi_0$, including inside continuation
constants.

Formally, these operations are defined by:
$$\begin{array}{r@{\quad}c@{\quad}l}
  x\{\diamond:=\pi_0\} &\equiv& x \\
  (\Lam{x}{t})\{\diamond:=\pi_0\} &\equiv&
  \Lam{x}{t\{\diamond:=\pi_0\}} \\
  (t\,u)\{\diamond:=\pi_0\} &\equiv&
  t\{\diamond:=\pi_0\}\,u\{\diamond:=\pi_0\} \\
  \kappa\{\diamond:=\pi_0\} &\equiv& \kappa \\
  (\k_{\pi})\{\diamond:=\pi_0\} &\equiv&
  \k_{(\pi\{\diamond:=\pi_0\})} \\[6pt]
  \diamond\{\diamond:=\pi_0\} &\equiv& \pi_0 \\
  (t\cdot\pi)\{\diamond:=\pi_0\} &\equiv&
  t\{\diamond:=\pi_0\}\cdot\pi\{\diamond:=\pi_0\} \\[6pt]
  (t\star\pi)\{\diamond:=\pi_0\} &\equiv&
  t\{\diamond:=\pi_0\}\star\pi\{\diamond:=\pi_0\} \\
\end{array}\eqno\begin{array}{r@{}}
  \\ \\ \\ (\kappa\in\K) \\ \\[6pt] \\ \\[6pt] \\
\end{array}$$
Note that when~$t$ is a proof-like term, we have
$t\{\diamond:=\pi_0\}\equiv t$.
From assumption~$(1)$ we immediately get:
\begin{lem}\label{l:StackExt}
  If $p\eval p'$, then
  $p\{\diamond:=\pi_0\}\eval p'\{\diamond:=\pi_0\}$ 
  (for all $\pi_0\in\Pi$).
\end{lem}
\proof
  By case analysis on the evaluation rule using~$(1)$.

(The same result holds if we replace $\eval$ by $\eval^*$.)

Let us now assume that~$t_0$ is a proof-like term (assumption~$(2)$)
that is a universal realizer of the formula $\existsN x\,f(x)=0$.
From Prop.~\ref{p:ExtrSigma01}, we know that there is some
$n\in\N$ such that $f(n)=0$ and
$$t_0\star(\Lam{xy}{y\,(\Stop\,x)})\cdot\diamond~\eval^*~
\Stop\star\widehat{n}\cdot\diamond\,.$$
But if we apply Lemma~\ref{l:StackExt} with an arbitrary stack~$\pi$,
we thus get
$$t_0\star(\Lam{xy}{y\,(\Stop\,x)})\cdot\pi~\eval^*~
\Stop\star\widehat{n}\cdot\pi$$
(using the fact that $t_0$ is a proof-like term, so that
$t_0\{\diamond:=\pi\}\equiv t_0$).

Since evaluation is deterministic and since the instruction~$\Stop$
has no evaluation rule, the answer produced by
Prop.~\ref{p:ExtrSigma01} with an arbitrary stack~$\pi$ is unique, and
it is the same as if we take the stack $\pi\equiv\diamond$.

\subsubsection{Adding other instructions}\
The property of independence of the witness w.r.t.\ the stack~$\pi$
crucially depends on the fact that evaluation is deterministic and
substitutive w.r.t.\ the stack constant~$\diamond$ (in the sense of
Lemma~\ref{l:StackExt}).
However, it is sometimes useful to consider instructions whose
evaluation rules break Lemma~\ref{l:StackExt} (without breaking
determinism of evaluation).
An example of such an instruction is the instruction $\Quote$ with the
evaluation rule
$$\Quote\star t\cdot\pi~\eval~t\star\widehat{n}_{\pi}\cdot\pi\,,$$
where $n_{\pi}$ is the \emph{code} of the stack~$\pi$ according to a
fixed bijection between natural numbers and stacks.
(Such an instruction is introduced in~\cite{Kri03} to realize several
forms of the axiom of choice.)
If~$t_0$ uses such an instruction, then the witness provided by
Prop.~\ref{p:ExtrSigma01} may actually depend on the stack~$\pi$.

\subsection{Extraction in the decidable case}
\label{ss:ExtrDec}

The witness extraction procedure we presented in
section~\ref{ss:ExtrSigma01} for $\Sigma^0_1$-formul{\ae} can be
generalized to any existential formula $\existsN x\,A(x)$ provided
the predicate $A(x)$ is decidable, using a decision function expressed
as a $\lambda_c$-term.

Formally, a \emph{decision function} for the predicate $A(x)$ is a
term $d_A\in\Lambda_c$ such that for all $n\in\N$,
$u,v\in\Lambda_c$ and $\pi\in\Pi$ we have
$$d_A\star\widehat{n}\cdot u\cdot v\cdot\pi
\quad\eval^*\quad\begin{cases}
  u\star\pi &\text{if}~\M\models A(n) \\
  v\star\pi &\text{if}~\M\not\models A(n) \\
\end{cases}$$
(writing~$\M$ the full standard model of PA2).
Intuitively, a decision function for the predicate $A(x)$ is a closed
$\lambda_c$-term~$d_A$ such that for every natural number
$n\in\N$, the applied term $d_A\,\widehat{n}$ acts as a boolean value
indicating whether the formula~$A(n)$ holds or not in the full
standard model of PA2.

Extracting a witness in this case also requires another
ingredient to repudiate the wrong witnesses proposed by the
realizer~$t_0$. 
Formally, we call a \emph{function of conditional refutation} of the
predicate $A(x)$ any term $r_A\in\Lambda_c$ such that
$$r_A~\urealK~\{n\}\limp\lnot A(n)$$
for all $n\in\N$ such that $\M\not\models A(n)$.
Intuitively, the purpose of a function of conditional refutation~$r_A$
is to provide a counter-realizer $t_A\,\widehat{n}\urealK\lnot A(n)$
that we shall oppose to the realizer $u\realK A(n)$ coming with any
wrong witness proposed by the realizer~$t_0$.
Such terms $r_A$ can be built for a very large class of formul{\ae}
as we shall see in section~\ref{ss:ExtrRefut}.

Using the decision function~$d_A$ and the function of conditional
refutation $r_A$, we now get a simple algorithm to perform witness
extraction from a universal realizer $t_0\urealK\existsN x\,A(x)$:
\begin{enumerate}[(1)]
\item Extract $n\in\N$ and $u\realK A(n)$ from the universal
  realizer~$t_0$.
\item Check whether $A(n)$ is true or not, using the decision
  function~$d_A$.
  \begin{enumerate}[$\bullet$]
  \item If $A(n)$ is true, then return $n$ (using the `$\Stop$'
    instruction).
  \item If $A(n)$ is false, then execute the realizer
    $r_A\,\widehat{n}\,u\realK\bot$ to backtrack.
  \end{enumerate}
\end{enumerate}
In the language~$\lambda_c$, this procedure is implemented by applying
the universal realizer~$t_0$ to the $\lambda_c$-term
$\Lam{xy}{d_A\,x\,(\Stop\,x)\,(r_A\,x\,y)}$ that does the expected
job:

\begin{prop}\label{p:ExtrDec}
  Let $d_A$ and~$r_A$ be respectively a decision function and
  a function of conditional refutation for the predicate $A(x)$,
  and let $t_0$ be a universal realizer of the formula
  $\existsN x\,A(x)$.
  Then for all $\pi\in\Pi$, the process
  $$t_0\star(\Lam{xy}{d_A\,x\,(\Stop\,x)\,(r_A\,x\,y)})\cdot\pi$$
  evaluates (in a finite number of steps) to a process of the form
  $\Stop\star\widehat{n}\cdot\pi$ for some natural number $n\in\N$
  such that $A(n)$ is true in the full standard model.
\end{prop}

\proof
  Let us take a stack~$\pi\in\Pi$ and work in the pole defined by
  $$\Bot~=~\{p~:~\exists n\,{\in}\,\N~~(\M\models A(n)
  ~~\text{and}~~p\eval^*\Stop\star\widehat{n}\cdot\pi)\}\,.$$
  Writing $S=\{\pi\}$, we easily check that
  $\Stop\realK\{n\}\limp\dot{S}$ for all $n\in\N$ such that
  $\M\models A(n)$
  (from the very definition of~$\Bot$ and~$S$).
  Let us now show that the term
  $\Lam{xy}{d_A\,x\,(\Stop\,x)\,(r_A\,x\,y)}$ realizes the
  formula $\forall x\,(\{x\}\limp A(x)\limp\dot{S})$.
  For that, consider an arbitrary element of the falsity value of this
  formula, that is: a stack of the form $\widehat{n}\cdot u\cdot\pi$
  for some $n\in\N$ and $u\in|A(n)|$.
  We have
  $$\Lam{xy}{d_A\,x\,(\Stop\,x)\,(r_A\,x\,y)}
  \star\widehat{n}\cdot u\cdot\pi ~\eval^*~
  d_A\star\widehat{n}\cdot(\Stop\,\widehat{n})\cdot
  (r_A\,\widehat{n}\,u)\cdot\pi\,.$$
  To show that $d_A\star\widehat{n}\cdot(\Stop\,\widehat{n})\cdot
  (r_A\,\widehat{n}\,u)\cdot\pi\in\Bot$,
  we distinguish two cases:
  \begin{enumerate}[$\bullet$]
  \item $\M\models A(n)$.\
    In this case we have
    $$d_A\star\widehat{n}\cdot(\Stop\,\widehat{n})\cdot
    (r_A\,\widehat{n}\,u)\cdot\pi~\eval^*~
    (\Stop\,\widehat{n})\star\pi~\eval~
    \Stop\star\widehat{n}\cdot\pi~\in~\Bot$$
    (using the fact that $\Stop\realK\{n\}\limp\dot{S}$ when
    $\M\models A(n)$), from which we get
    $d_A\star\widehat{n}\cdot(\Stop\,\widehat{n})\cdot
    (r_A\,\widehat{n}\,u)\cdot\pi\in\Bot$
    by anti-evaluation.
  \item $\M\not\models A(n)$.\
    In this case we have
    $$d_A\star\widehat{n}\cdot(\Stop\,\widehat{n})\cdot
    (r_A\,\widehat{n}\,u)\cdot\pi~\eval^*~
    (r_A\,\widehat{n}\,u)\star\pi~\eval^*~
    r_A\star\widehat{n}\cdot u\cdot\pi~\in~\Bot$$
    since
    $\widehat{n}\cdot u\cdot\pi\in\|\{n\}\limp A(n)\limp\bot\|$ and
    $r_A\in|\{n\}\limp A(n)\limp\bot|$ from the definition that~$r_A$
    is a function of conditional refutation.
    By anti-evaluation we get:
    $d_A\star\widehat{n}\cdot(\Stop\,\widehat{n})\cdot
    (r_A\,\widehat{n}\,u)\cdot\pi\in\Bot$.
  \end{enumerate}
  In both cases we deduce that
  $\Lam{xy}{d_A\,x\,(\Stop\,x)\,(r_A\,x\,y)}
  \star\widehat{n}\cdot u\cdot\pi\in\Bot$ by
  anti-evaluation, which finishes the proof that
  $\Lam{xy}{d_A\,x\,(\Stop\,x)\,(r_A\,x\,y)}$ realizes the formula
  $\forall x\,(\{x\}\limp A(x)\limp\dot{S})$ in the pole~$\Bot$.
  From the latter, we immediately deduce that
  $(\Lam{xy}{d_A\,x\,(\Stop\,x)\,(r_A\,x\,y)})\cdot\pi\in
  \|\existsN x\,f(x)=0\|$, from which we conclude that
  $t_0\star(\Lam{xy}{d_A\,x\,(\Stop\,x)\,(r_A\,x\,y)})\cdot\pi
  \in\Bot$.\qed

\subsubsection{The particular case of $\Sigma^0_1$-formul{\ae}}\
In the case where the predicate $A(x)$ is of the form
$A(x)\equiv f(x)=0$ for some primitive recursive function symbol~$f$,
it is easy to implement a decision function~$d_A$ from a
$\lambda_c$-term that actually computes~$f$.
Such a function~$d$ that tests whether $f(n)=0$ for a given argument
$n\in\N$ can even be characterized in terms of realizability as follows:
\begin{lem}
  Let~$f$ be a primitive recursive function symbol.
  For every term $d\in\Lambda_c$, the following assertions are
  equivalent:
  \begin{enumerate}[\em(1)]
  \item $d$ decides the predicate $A(x)\equiv f(x)=0$;
  \item $d\urealK\forall Z\,\forallN x\,\bigl(Z(0)\limp
    \forall y\,Z(s(y))\limp Z(f(x))\bigr)$.
  \end{enumerate}
\end{lem}

\proof
  $1.\limp 2.$\ Easily follows from the evaluation rules of the
  term~$d$.\par\noindent
  $2.\limp 1.$\
  Let $n\in\N$, $u,v\in\Lambda_c$ and $\pi\in\Pi$.
  We distinguish two cases:
  \begin{enumerate}[$\bullet$]
  \item $f(n)=0$.\
    We let $\Bot=\{p~\eval^*~u\star\pi\}$ and define a function
    $F:\N\to\P(\pi)$ by $F(0)=\{\pi\}$ and $F(p)=\varnothing$ for all
    $p>0$.
    We easily check that $u\in|\dot{F}(0)|$,
    $v\in|\forall y\,\dot{F}(s(y))|$ and $\pi\in\|\dot{F}(f(n))\|$,
    hence $d\star\widehat{n}\cdot u\cdot v\cdot\pi\in\Bot$.
  \item $f(n)\neq 0$.\
    We let $\Bot=\{p~\eval^*~v\star\pi\}$ and define a function
    $F:\N\to\P(\pi)$ by $F(0)=\varnothing$ and $F(p)=\{\pi\}$ for all
    $p>0$.
    We again check that $u\in|\dot{F}(0)|$,
    $v\in|\forall y\,\dot{F}(s(y))|$ and $\pi\in\|\dot{F}(f(n))\|$,
    hence $d\star\widehat{n}\cdot u\cdot v\cdot\pi\in\Bot$.\qed
  \end{enumerate}

The function of conditional refutation for the predicate
$A(x)\equiv f(x)=0$ is even easier to build: simply take the constant
function $r_A\equiv\Lam{\_z}{z\,?}$ (where~$?$ is any $\lambda_c$-term
possibly depending on~$z$), using the fact that
$\TYPK{}{\Lam{z}{z\,?}}{f(n)\neq 0}$
for all natural numbers~$n$ such that $f(n)\neq 0$.
(Note that the term $\Lam{z}{z\,?}$ does not depend on~$n$.)
In this case, the function of conditional refutation~$r_A$
can be replaced by the \emph{conditional refutation} 
$\Lam{z}{z\,?}$ that is a universal realizer of the formula
$f(n)\neq 0$ for all natural numbers~$n$ such that $f(n)\neq 0$.

Given a term $d_f\in\Lambda_c$ that decides the predicate $f(x)=0$, we
can thus perform witness extraction from a universal realizer
$t_0\urealK\existsN x\,f(x)=0$ using the process:
$$t_0\star(\Lam{xy}{d_f\,x\,(\Stop\,x)\,(y\,?)})\star\pi$$
(whose second branch has been simplified).
Note that this process is slightly more complex than the process
presented in Prop.~\ref{p:ExtrSigma01} that does not even need to
consider a decision function to perform witness extraction
from~$t_0$.

\subsection{Existence of functions of conditional refutation}
\label{ss:ExtrRefut}

The existence of a function of conditional refutation can be shown for
a wide class of predicates, and in particular for every predicate
$A(x)$ that is expressed in the language of first-order arithmetic
such as defined in the end of section~\ref{ss:Degenerate}
(replacing $\forall x\,(\Nat(x)\limp P)$ by $\forallN x\,P$ in the
corresponding BNF).

Let us first recall that:
\begin{prop}\label{p:PrenexRealiz}
  For every $k\ge 0$, there exists a closed proof-term $R_k$ such
  that for every formula of the form
  $$A~\equiv~\existsN y_1\,\forallN z_1\cdots
  \existsN y_k\,\forallN z_k~f(y_1,z_1,\ldots,y_k,z_k)\neq 0\,,$$
  if $\M\models A$, then $R_k\urealK A$.
\end{prop}

\proof
  The existence of such a proof-term $R_k$ is an immediate consequence
  of Theorem~21 (p.~14) in~\cite{Kri05}.
  Note that $R_k$ only depends on~$k$.\qed

We also check that:
\begin{prop}[Existence of the prenex form]\label{p:PrenexEx}
   If $A(x_1,\ldots,x_p)$ is a formula of first-order arithmetic
  depending on~$p$ first-order variables~$x_1,\ldots,x_p$,
  then there exists a natural number $k\ge 0$ and a function
  symbol~$f$ of arity $p+2k$ such that the formula
  $$A'(x_1,\ldots,x_p)~\equiv~\existsN y_1\,\forallN z_1\cdots
  \existsN y_k\,\forallN z_k~
  f(x_1,\ldots,x_p,y_1,z_1,\ldots,y_k,z_k)\neq 0$$
  is logically equivalent to $A(x_1,\ldots,x_p)$, in the sense that
  there are closed proof-terms $u_1,u_2$ such that
  $$\begin{array}{l}
    \TYPK{}{~u_1~}{~\forallN x_1\cdots\forallN x_p\,
      (A(x_1,\ldots,x_p)\limp A'(x_1,\ldots,x_p))} \\
    \TYPK{}{~u_2~}{~\forallN x_1\cdots\forallN x_p\,
      (A'(x_1,\ldots,x_p)\limp A(x_1,\ldots,x_p))}\,. \\
  \end{array}$$
\end{prop}

\proof
  This theorem is the reformulation in the type system of
  Fig.~\ref{f:PA2} and~\ref{f:PA2ext} of the existence of prenex
  forms in first-order arithmetic.\qed

From Prop.~\ref{p:PrenexRealiz} and~\ref{p:PrenexEx} we deduce the
following:
\begin{prop}[Existence of a conditional refutation]
  If $A(x)$ is a formula of first-order arithmetic that only
  depends on a first-order variable~$x$, then the predicate
  $A(x)$ has a function of conditional refutation~$r_A$.
\end{prop}

\proof
  From Prop.~\ref{p:PrenexEx}, there exists a formula
  $$A'(x)~\equiv~\existsN y_1\,\forallN z_1\cdots
  \existsN y_k\,\forallN z_k~f(x,y_1,z_1,\ldots,y_k,z_k)\neq 0$$
  with closed proof-terms $u_1,u_2$ such that:
  $$\TYPK{}{~u_1~}{~\forallN x\,(A(x)\limp A'(x))}
  \qquad\text{and}\qquad
  \TYPK{}{~u_2~}{~\forallN x\,(A'(x)\limp A(x))}\,.$$
  It suffices to take $r_A\equiv\Lam{x}{u_2\,x\,R_k}$
  (by Prop.~\ref{p:PrenexRealiz}).\qed

\subsection{The method of the kamikaze}
\label{ss:ExtrKam}

The witness extraction procedure presented in section~\ref{ss:ExtrDec}
depends on two components: a function~$d_A$ deciding the predicate
$A(x)$, and another function~$r_A$ that conditionally refutes the
predicate $A(x)$.
The critical component here is the decision function~$d_A$, since the
function~$r_A$ can be constructed for a wider class of formul{\ae},
i.e. for all arithmetic formul{\ae} (cf section~\ref{ss:ExtrRefut}).

In the case where we have a function of conditional refutation~$r_A$
but no decision function for the predicate $A(x)$---typically,
when~$A(x)$ is a non atomic arithmetic formula---we can still extract
a possibly infinite sequence of `witness proposals' from the
universal realizer $t_0\urealK\forallN x\,A(x)$ by systematically
repudiating every proposed witness using the function of conditional
refutation~$r_A$.

This extraction method, which we call the \emph{method of the
  kamikaze}, consists to apply the universal
realizer~$t_0\urealK\existsN x\,A(x)$ to the term
$\Lam{xy}{\Print\,x\,(r_A\,x\,y)}$ (using the `$\Print$' instruction
introduced in Remark~\ref{r:Print}), thus implementing in the
language~$\lambda_c$ the following algorithm:
\begin{enumerate}[(1)]
\item Extract $n\in\N$ and $u\realK A(n)$ from the universal
  realizer~$t_0$.
\item Print~$n$ on some output device.
\item Try to backtrack by executing $r_A\,\widehat{n}\,u$.
\end{enumerate}
The crucial point here is that there is no warranty that the piece of
code executed at step~3 will actually issue a backtrack, since we do
not know whether $\lnot A(n)$ is true.
The only invariant we can ensure is the following:
as long as the proposed witness~$n$ is incorrect, the refutation
function~$r_A$ is applied in agreement with its specification, so that
step~3.\ will issue a backtrack.
But as soon as a correct witness~$n$ has been reached, the current
process becomes ill-typed, and then anything may happen:
the process may enter an infinite loop (possibly displaying other
numbers) as it may crash, for instance due to a stack underflow (by
evaluating an abstraction or one of the instructions~$\cc$,
$\k_{\pi}$, $\Print$ in front of an empty stack), or due to the fact
that~$\Print$ is evaluated in front of a stack which does not start
with a primitive numeral.

Of course, the interest of the method is that the process that
performs the blind extraction of the successive witnesses proposed by
the universal realizer~$t_0$ cannot go wrong until a correct witness
has been reached.
We can actually even show that this process eventually reaches a
correct witness:

\begin{prop}\label{p:ExtrKam}
  If $t_0$ is a universal realizer of $\existsN x\,A(x)$ and
  if~$r_A$ is a function of conditional refutation of the predicate
  $A(x)$, then for all stacks $\pi\in\Pi$ the process
  $$t_0\star(\Lam{xy}{\mathsf{print}\,x\,(r_A\,x\,y)})\cdot\pi$$
  evaluates (in a finite number of steps) to a process of the form
  $\mathsf{print}\star\widehat{n}\cdot u\cdot\pi$, where
  $u\in\Lambda_c$ and where $n\in\N$ is such that $A(n)$ holds in the
  full standard model.
\end{prop}

\proof
  Let us take a stack~$\pi\in\Pi$ and work in the pole defined by
  $$\Bot~=~\{p~:~\exists n\,{\in}\,\N~\exists u\,{\in}\,\Lambda_c~
  (\M\models A(n)~\land~
  p\eval^*\mathsf{print}\star\bar{n}\cdot u\cdot\pi)\}\,.$$
  Set $S=\{\pi\}$.
  We first want to show that $\mathsf{print}\star
  \widehat{n}\cdot(r_A\,\widehat{n}\,v)\cdot\pi\in\Bot$
  for all $n\in\N$ and for all $v\in|A(n)|$.
  We distinguish the following two cases:
  \begin{enumerate}[$\bullet$]
  \item $\M\models A(n)$.\
    In this case we have
    $\mathsf{print}\star
    \widehat{n}\cdot(r_A\,\widehat{n}\,v)\cdot\pi\in\Bot$
    from the very definition of the pole~$\Bot$.
  \item $\M\not\models A(n)$.\
    In this case, we have
    $$\mathsf{print}\star
    \widehat{n}\cdot(r_A\,\widehat{n}\,v)\cdot\pi
    ~\eval~(r_A\,\widehat{n}\,v)\star\pi~\eval^*~
    r_A\star\widehat{n}\cdot v\cdot\pi~\in~\Bot$$
    from our assumption on~$r_A$ combined with the fact that
    $\M\not\models A(n)$.
    Hence we get
    $\mathsf{print}\star
    \widehat{n}\cdot(r_A\,\widehat{n}\,v)\cdot\pi\in\Bot$ by
    anti-evaluation.
  \end{enumerate}
  From this result we easily get
  $$\Lam{xy}{\mathsf{print}\,x\,(r_A\,x\,y)}~\realK~
  \forall x\,(\{x\}\limp A(x)\limp\dot{S})$$
  and finally:\
  $t_0\star(M\,\Lam{xy}{\mathsf{print}\,x\,(r_A\,x\,y)})\cdot\pi
  \in\Bot$.\qed

\medbreak
Let us note that the above proof relies in an essential way in the
definition of a pole~$\Bot$ that is not closed under evaluation, thus
reflecting the fact that the process which performs kamikaze
extraction is correct \emph{up to some point} during evaluation.
After this point has been reached---that is: when a correct witness
has been printed---the realizability model gives us no invariant
anymore about the execution of the current process, so that anything
may happen.

\COUIC{
\medbreak
In the sequel, we will focus on the witness extraction method
described in section~\ref{ss:ExtrSigma01}, by analyzing it
through a suitable negative translation.
}

\section{An example based on the minimum principle}
\label{s:Example}
\def\Minus{\mathsf{minus}}
\def\FUN{\mathsf{Fun}}
\def\MINPRINC{\mathsf{MinPrinc}}
\def\TestLe{\mathsf{test\_le}}
\def\MinPrinc{\mathsf{min\_princ}}
\def\MinAux{\mathsf{min\_aux}}

\noindent In this section, we give an example of witness extraction in the
$\Sigma^0_1$-case.

An important aspect of the witness extraction procedure described in
Prop.~\ref{p:ExtrSigma01} is that the universal realizer
$t_0\urealK\existsN x\,f(x)=0$ does not need to be a proof-term in the
sense of the type system of~$\PAext$---it just needs to be a universal
realizer in the sense of classical realizability.
Indeed, the naive method that consists to extract the $\lambda_c$-term
from the proof \emph{as is} tends to produce highly inefficient code.
On the other hand, many useful arithmetic lemmas have universal
realizers that are much more compact (and much more efficient) than
the realizers that would come from official proofs.

For this reason, it is reasonable to isolate such lemmas during the
extraction process, and to replace their official proof-terms (i.e.\
coming from derivations in~$\PAext$) by universal realizers built by
hand.
In what follows, we shall illustrate this point with the minimum
principle.

\subsection{Notations}

In $\PAext$, it is convenient to define the ordering relation $x\le y$
from Leibniz equality by letting
$$x\le y~\equiv~\Minus(x,y)=0\,,$$
where~$\Minus$ is the binary primitive recursive function defined by
the equations
$$\begin{array}{r@{~~}c@{~~}l}
  \Minus(x,0) &=& x \\
  \Minus(0,s(y)) &=& 0 \\
  \Minus(s(x),s(y)) &=& \Minus(x,y) \\
\end{array}$$

Given a unary primitive recursive function symbol~$f$, we express
that~$f$ is a function from natural numbers to natural numbers with
the formula
$$\FUN(f)~\equiv~\forallN x\,\Nat'(f(x))
~\equiv~\forall x\,(\{x\}\limp
\forall Z\,((\{f(x)\}\limp Z)\limp Z))\,.$$
It is easy to check that universal realizers of the formula~$\FUN(f)$
are precisely the closed $\lambda_c$-terms that compute the
function~$f$, namely:
\begin{lem}
  Given a term~$t\in\Lambda_c$, the following assertions are
  equivalent:
  \begin{enumerate}[\em(1)]
  \item For all $u\in\Lambda_c$, $\pi\in\Pi$:\quad
    $t\star\widehat{n}\cdot u\cdot\pi\succ^*
    u\star\widehat{f(n)}\cdot\pi$\hfill
    (i.e. $t$ computes~$f$)
  \item $t\urealK\FUN(f)$\hfill
    (i.e.\ $t$ universally realizes $\FUN(f)$)
  \end{enumerate}
\end{lem}

\proof
  1. $\limp$ 2.\ immediately follows from the definitions of classical
  realizability.\par\noindent
  2. $\limp$ 1.\ Let us assume that $t\urealK\FUN(f)$, and
  fix~$n\in\N$, $u\in\Lambda_c$ and $\pi\in\Pi$.
  We define the pole
  $\Bot=\{p~:~p\succ^*u\star\widehat{f(n)}\cdot\pi\}$
  and the falsity value $S=\{\pi\}$, from which we easily check that
  $u\realK\{f(n)\}\limp\dot{S}$.
  From our initial assumption, we have
  $t\urealK\{n\}\limp(\{f(n)\}\limp\dot{S})\limp\dot{S}$,
  and thus $t\star\widehat{n}\cdot u\cdot\pi\in\Bot$.\qed

Finally, we use the shorthand $\<x;y\>\equiv\Lam{z}{z\,x\,y}$ to
denote order pairs in~$\lambda_c$, keeping in mind that this
construction can be used to prove (or realize) both conjunctions and
numeric existential quantifications.

\subsection{The functional minimum principle}

We now want to build a universal realizer of the formula expressing
that a function from natural numbers to natural numbers reaches its
minimum:
$$\MINPRINC~~\equiv~~
\FUN(f)\limp\existsN x\,\forallN y\,(f(x)\le f(y))$$
(Note that the premise $\FUN(f)$ is crucial to prove$/$realize
the result.)
Since this formulation of the minimum principle is (classically)
provable in~$\PAext$, we could take any proof-term of it as a
universal realizer.
In this case however, it is much more interesting to build a universal
realizer by hand.

For that, let us take a closed $\lambda_c$-term $\TestLe$ that
performs the comparison of two primitive natural numbers, in the sense
that
$$\TestLe\star\widehat{n}\cdot\widehat{m}\cdot u\cdot v\cdot\pi
  ~~\succ^*~~\begin{cases}
    u\star\pi &\text{if}~n\le m\\
    v\star\pi &\text{otherwise} \\
\end{cases}$$
for all $n,m\in\N$, $u,v\in\Lambda_c$ and $\pi\in\Pi$.
(It is a straightforward exercise of programming to implement such a
term in~$\lambda_c$.)

Now, let us consider a closed $\lambda_c$-term $\MinAux$ such that
$$\begin{array}{l}
  \MinAux\star f\cdot k\cdot n\cdot m\cdot\pi~~\succ^* \\
  \qquad\<n,\Lam{n'}{f\,n'\,(\Lam{m'}{
      \TestLe\,m\,m'\,\mathbf{I}\,
      (k~(\MinAux\,f\,k\,n'\,m'))})}\>\cdot\pi \\
\end{array}$$
for all $f,k,n,m\in\Lambda_c$ and $\pi\in\Pi$.
Intuitively, such a $\lambda_c$-term $\MinAux$ is a recursive function
that takes the following arguments:
\begin{enumerate}[$\bullet$]
\item A realizer $f\realK\FUN(f)$
  (i.e. an implementation of~$f$)
\item A continuation $k\realK
  \lnot\existsN x\,\forallN y\,(f(x)\le f(y))$ for backtracking.
\item The current witness proposal~$n$.
\item The image $m=f(n)$ of the current witness proposal.
  (We keep this argument across the recursive call to avoid
  recomputing it later.)
\end{enumerate}
When it is called with these arguments, the function $\MinAux$ returns
an ordered pair $\<n,h\>$ whose first component is the current witness
proposal~$n$, and whose second component is a function 
$$h~\equiv~\Lam{n'}{f\,n'\,(\Lam{m'}{
      \TestLe\,m\,m'\,\mathbf{I}\,
      (k~(\MinAux\,f\,k\,n'\,m'))})}$$
that takes a natural number $n'$, computes its image $m'=f(n')$ and
compares it with~$m$.
In the case where $m\le m'$, the function~$h$ returns the identity
term~$\mathbf{I}$, which is an obvious realizer of $f(n)\le f(n')$.
In the case where $m'<m$, the function~$f$ backtracks using the
continuation~$k$, and recursively calls~$\MinAux$ with~$n'$ as the new
witness proposal, and~$m'$ as its image by~$f$.

Note that there are several ways to implement the term~$\MinAux$
in~$\lambda_c$.
For instance, we can let
$$\MinAux~\equiv~\mathbf{Y}\,(\Lam{rfknm}{
  \<n,\Lam{n'}{f\,n'\,(\Lam{m'}{
      \TestLe\,m\,m'\,\mathbf{I}\,
      (k~(r\,f\,k\,n'\,m'))})}\>
})\,,$$
where $\mathbf{Y}\equiv(\Lam{yz}{z(yy)})(\Lam{yz}{z(yy)})$
is Turing's fixpoint combinator%
\footnote{We using Turing's fixpoint combinator rather than Church's,
  since Turing's combinator is better suited for the call-by-name
  strategy.};
or we can simply introduce $\MinAux$ as an extra instruction with the
desired evaluation rule.
Whatever the way we implement $\MinAux$, we can check that:
\begin{lem}\label{l:MinAux}
  Writing $E\equiv\existsN x\,\forallN y\,f(x)\le f(y)$,
  we have:
  $$\MinAux~~\urealK~~\forall x\,
  (\FUN(f)\limp\lnot E\limp\{x\}\limp\{f(x)\}\limp E)\,.$$
\end{lem}

\proof
  Fix a pole~$\Bot$ and two realizers $f\realK\FUN(f)$ and
  $k\realK\lnot E$, and consider the property $\text{IH}(m)$
  defined by 
  $$\begin{array}{l@{\quad}l}
    \text{IH}(m):& \text{for all $n\in\N$ s.t. $f(n)=m$,
      for all $\pi\in\|E\|$} \\
    &\text{we have:}\quad \MinAux\star f\cdot
    k\cdot\widehat{n}\cdot\widehat{m}\cdot\pi\in\Bot\,. \\
  \end{array}$$
  We want to prove $\text{IH}(m)$ by well-founded induction on~$m$.
  For that, let us fix $m\in\N$, assume that $\text{IH}(m')$ for
  all $m'<m$, and take $n\in\N$ such that $f(n)=m$ and $\pi\in\|E\|$.
  From the evaluation rules of~$\TestLe$, we can derive that
  $$\TestLe~\widehat{m}~\widehat{m'}~
  \mathbf{I}~(k\,(\MinAux\,f\,k\,n'\,m'))
  ~\realK~f(n)\le f(n')$$
  for all $n',m'\in\N$ such that $m'=f(n')$, distinguishing cases
  depending on whether $m\le m'$ or $m'<m$, and using the induction
  hypothesis $\text{IH}(m')$ in the second case.
  From this we successively get
  $$\begin{array}{l}
    \Lam{m'}{\TestLe~\widehat{m}~\widehat{m'}~
      \mathbf{I}~(k\,(\MinAux\,f\,k\,n'\,m'))}~\realK~
    \forall y\,(\{f(y)\}\limp f(n)\le f(y)) \\
    \Lam{n'}{f\,n'\,(\Lam{m'}{
        \TestLe~\widehat{m}~\widehat{m'}~
        \mathbf{I}~(k\,(\MinAux\,f\,k\,n'\,m'))})}~\realK~
    \forallN y\,(f(n)\le f(y)) \\
    \<n,\Lam{n'}{f\,n'\,(\Lam{m'}{
        \TestLe~\widehat{m}~\widehat{m'}~
        \mathbf{I}~(k\,(\MinAux\,f\,k\,n'\,m'))})}\>~\realK~E \\
    \<n,\Lam{n'}{f\,n'\,(\Lam{m'}{
        \TestLe~\widehat{m}~\widehat{m'}~
        \mathbf{I}~(k\,(\MinAux\,f\,k\,n'\,m'))})}\>\star\pi
    ~~\in~~\Bot \\
  \end{array}$$
  hence
  $\MinAux\star f\cdot
  k\cdot\widehat{n}\cdot\widehat{m}\cdot\pi\in\Bot$,
  by anti-evaluation.\qed

Now we can set:
$$\MinPrinc~\equiv~\Lam{f}{f\,\widehat{0}\,(\Lam{m}{\cc\,
    (\Lam{k}{\MinAux\,f\,k\,\widehat{0}\,m})})}\,.$$
Intuitively, this function takes an implementation of~$f$, computes
the image $m=f(0)$, captures the current continuation as~$k$ and then
calls $\MinAux$ with $0$ as the initial witness proposal (accompanied
with its image $m=f(0)$).

Combining Lemma~\ref{l:MinAux} and the property of adequacy
(Prop.~\ref{p:KAdequacy}) with the derivable judgment
$$\begin{array}{l}
  z:\forall x\,(\FUN(f)\limp\lnot E\limp\{x\}\limp\{f(x)\}\limp E)\\
  {\vdash_{\mathrm{NK}}}~~
  \Lam{f}{f\,\widehat{0}\,(\Lam{m}{\cc\,
      (\Lam{k}{z\,f\,k\,\widehat{0}\,m})})}~~:~~\MINPRINC \\
\end{array}$$
we immediately deduce that:
$$\MinPrinc~~\urealK~~\FUN(f)\limp
\existsN x\,\forallN y\,(f(x)\le f(y))\,.$$

\subsection{A $\Sigma^0_1$-consequence of the minimum principle}

Let~$f$ and~$g$ be two functions from natural numbers to natural
numbers.
The minimum principle gives a simple argument to show the existence of
a natural number~$x$ such that $f(x)\le f(g(x))$, which is to take a
point~$x$ where~$f$ reaches its minimum.
In~$\PAext$, the argument is formalized as follows:
$$\begin{array}{l}
  z:\MINPRINC,~f:\FUN(f),~g:\FUN(g)~~{\vdash}_{\text{NK}} \\
  \qquad z\,f\,(\Lam{nh}{\<n,\check{g}\,h\>})
  ~~:~~\existsN x\,(f(x)\le f(g(x))) \\
\end{array}$$

Considering implementations $\check{f}\urealK\FUN(f)$ and
$\check{g}\urealK\FUN(g)$ of the functions~$f$ and~$g$, we thus get a
universal realizer of the following $\Sigma^0_1$-formula:
$$\MinPrinc\,\check{f}\,(\Lam{nh}{\<n,\check{g}\,n\,h\>})
~~\urealK~~\existsN x\,(f(x)\le f(g(x)))$$
By Prop.~\ref{p:ExtrSigma01}, we know that the process
$$p_0~~\equiv~~
\MinPrinc\,\check{f}\,(\Lam{nh}{\<n,\check{g}\,n\,h\>})
\star(\Lam{xy}{y\,(\Stop\,x)})\cdot\diamond$$
computes the desired witness (which depends of course on~$f$
and~$g$).

\subsection{Executing $\lambda_c$-code}

Fig.~\ref{f:ExecTrace} illustrates the execution of the above
process~$p_0$ in the particular case where~$f$ and~$g$ are given by
$$f(x)=|x-1000|\qquad\text{and}\qquad g(x)=2x+1\,.$$
The process~$p_0$ was executed using the \texttt{jivaro} head
reduction machine~\cite{Jiv09},
a small interpretor of Krivine's $\lambda_c$-calculus extended with
many built-in primitives (mainly for arbitrary-precision arithmetic and
string manipulation).
We slightly altered the code of~$p_0$ in order to print intermediate
witness proposals, so that the actual code of~$p_0$ is
$$p_0~~\equiv~~
\MinPrinc\,\check{f}\,(\Lam{nh}{\<n,\check{g}\,n\,h\>})
\star(\Lam{xy}{\Print\,x\,y\,(\Stop\,x)})\cdot\diamond$$
where~$\Print$ is the instruction mentioned in Remark~\ref{r:Print}
p.~\pageref{r:Print}.
\begin{figure}[htp]
  \begin{center}
    \begin{minipage}{110mm}
      \begin{center}
        \hrule\vspace{1.5pt}\hrule\medbreak
        \underline{\textbf{Input script}}\medbreak
        \begin{minipage}{102mm}
          \scriptsize
\begin{verbatim}
Define I x = x ;;                      (* Identity *)
Define pair x y z = z x y ;;           (* Pairing *)
Define test_le = int_le ;;             (* Alias for int_le primitive *)

(* Realizing the minimum principle *)
Define min_aux f k n m = pair n (min_snd f k m) ;;
Define min_snd f k m n' = f n' (\m' test_le m m' I (k (min_aux f k n' m'))) ;;
Define min_princ f = f 0 (\n (callcc (\k min_aux f k 0 n))) ;;

(* Take   f(x) = |n - 1000|   and   g(x) = 2x + 1 *)
Define f n = int_le n 1000 (int_minus 1000 n) (int_minus n 1000) ;;
Define g n = int_mult 2 n (\m int_succ m) ;;

(* Universal realizer of \existsN x, f(x) <= f(g(x)) *)
Define realizer = min_princ f (\n\h pair n (g n h)) ;;

Trace On ;;
(* Perform Sigma^0_1 witness extraction & print intermediate witnesses *)
Eval realizer ; (\x\y print x y (stop x)) ;;
\end{verbatim}
        \end{minipage}\bigbreak
        \begin{minipage}[t]{32mm}
          \centerline{\underline{\textbf{Output}}}\bigbreak
          \scriptsize
\begin{verbatim}
0
1
3
7
15
31
63
127
255
511
1023
0.01 s: stopped
Final state: stop ; 1023
\end{verbatim}
        \end{minipage}\quad
        \begin{minipage}[t]{70mm}
          \centerline{\underline{\textbf{Evaluation statistics
              (instruction calls)}}}
          \medbreak\footnotesize
          \vrule\quad\begin{tabular}{l@{\qquad}r}
            $@$ (\textsc{Push}) & 419 \\
            $\lambda$ (\textsc{Grab}) & 68 \\ \\
            \texttt{int\_le} & 23 \\
            \texttt{pair} & 22 \\
            \texttt{f} & 12 \\
            \texttt{int\_minus} & 12 \\
            \texttt{g} & 11 \\
            \texttt{int\_mult} & 11 \\
            \texttt{int\_succ} & 11 \\
          \end{tabular}\quad\vrule\quad
          \begin{tabular}{l@{\qquad}r}
            \texttt{min\_aux} & 11 \\
            \texttt{min\_snd} & 11 \\
            \texttt{print} & 11 \\
            \texttt{test\_le} & 11 \\
            $\k_{\pi}$ (\textsc{Restore}) & 10 \\
            \texttt{I} & 1 \\
            \texttt{callcc} (\textsc{Save}) & 1 \\
            \texttt{min\_princ} & 1 \\
            \texttt{realizer} & 1 \\
            \texttt{stop} & 1 \\
          \end{tabular}
        \end{minipage}
        \medbreak\hrule\vspace{1.5pt}\hrule
      \end{center}
    \end{minipage}
  \end{center}\vspace{-12pt}
  \caption{Example of witness extraction using the
    \texttt{jivaro} machine}
  \label{f:ExecTrace}
\end{figure}

As shown in the input script of Fig.~\ref{f:ExecTrace}, each component
of the process~$p_0$ is introduced as a new instruction given with its
evaluation rule (using the command \texttt{Define}).
Note that such definitions may be (mutually) recursive, which is the
case here for the instructions \texttt{min\_aux} and
\texttt{min\_snd}.
The interest of using named instructions rather than anonymous
$\lambda_c$-terms is that we can more easily track when each piece
of the code comes into head position during execution.

The output given in Fig.~\ref{f:ExecTrace} shows that during its
execution, the process~$p_0$ successively tries the following guesses
for~$x$:
$$\begin{array}{l}
  x_0=0,\quad x_1=1,\quad x_2=3,\quad x_3=7,\quad x_5=15,\quad
  x_6=31,\\
  x_7=63,\quad x_8=127,\quad x_9=255,\quad x_{10}=511,\quad
  x_{11}=1023. \\
\end{array}$$
Since the last guess ($x_{11}=1023$) is a solution of the problem, the
execution stops on the final state
$\Stop\star\widehat{1023}\cdot\diamond$,
with the form predicted by Prop.~\ref{p:ExtrSigma01}.

The choice of this particular sequence of guesses is explained as
follows.

During the execution of the process~$p_0$, the proof of
$\existsN x\,(f(x)\le f(g(x)))$ uses the guess~$x_i$
produced by the minimum principle as well as the accompanying
justification of the formula $\forallN y\,(f(x_i)\le f(y))$ to build a
realizer of $f(x_i)\le f(g(x))$.
But when the latter is executed, it invokes the accompanying
justification, that actually compares the values of $x_i$
and $g(x_i)$ by~$f$.
In the case where $f(g(x_i))<f(x_i)$, the guess~$x_i$ was wrong, and
the accompanying justification backtracks to the point where the
minimum principle was invoked (using an embedded continuation
$\k_{\pi}$).
When restarted, the minimum principle can then propose
$x_{i+1}=g(x_i)$ as a new guess.
As a consequence, the process~$p_0$ produces its guesses $x_i=g^i(0)$
by iterating the function~$g$ until $f(x_i)\le f(g(x_i))$.%
\footnote{Note that although each guess $x_i$ claims to be a point
  where~$f$ reaches its minimum (until the context proves it wrong and
  forces backtrack), none of them---including the last one---is such a
  point, since~$f$ takes its minimum for $x=1000$.}

Note that this behavior is the same as the one we observe when
treating the same example using Friedman's method or its
refinements~\cite{BBS02}.
Of course, this similarity is not a coincidence since Friedman's
translation is actually hard-wired in Krivine's semantics (as already
pointed out in~\cite{Oli08}), and we shall come back to this point
with more details in sections~\ref{s:HA2} and~\ref{s:NegTrans}.

\subsubsection{Evaluation statistics}\
Fig.~\ref{f:ExecTrace} also provides some statistics giving how many
times each instruction has been called during evaluation.

Not surprisingly, the most frequent operations are
\textsc{Push} (419 times) and \textsc{Grab} (68 times),
the asymmetry between these coming from the fact that stack arguments
are not only consumed by abstractions (\textsc{Grab}), but also by the
instructions used by the program, which may be primitive
(\texttt{callcc}, \texttt{int\_le}, etc.) or defined by the user
(\texttt{pair}, \texttt{min\_aux}, etc.)

We can also see that our hand-made implementation of the minimum
principle is optimal: the number of calls to the function~\texttt{f}
as well as the number of comparisons of images (by~$f$) of guesses
(using the instruction \texttt{test\_le}) are both minimal.
Moreover, the \texttt{callcc} instruction is called once during the
whole execution, thus creating a unique continuation constant
$\k_{\pi}$ (where $|\pi|=2$) that is used exactly 10 times
(\textsc{Restore}), that is: once for each backtrack.

We also tested this example by replacing the hand-made realizer of the
minimum principle with an actual proof of it (in~$\PAext$).
The observed behavior remains the same, but the proof-term is much
bigger and its execution is quite inefficient, mainly due to the
arithmetic reasoning involved in the induction underlying the proof of
the principle.
(In the hand-made realizer, induction is performed at the
meta-theoretic level, and thus has no cost during execution.)
We can also notice that depending the way we use classical logic in
the proof of the minimum principle, the corresponding proof-term may
invoke several times the call$/$cc instruction, or only once as in the
hand-made realizer.

\section{Intuitionistic second-order arithmetic}
\label{s:HA2}

\noindent We now define a type system for intuitionistic second-order
arithmetic (HA2), as well as a realizability model that closely
follows the traditional Brouwer-Heyting-Kolmogorov interpretation.
As in~\cite{Oli08}, we introduce a primitive form of conjunction (as a
Cartesian product) and primitive forms of first- and second-order
existential quantification (as infinitary unions).

\subsection{The language of formul{\ae}}
\label{ss:LangHA2}

The language of arithmetic expressions of HA2 is the same as for PA2
(Fig.~\ref{f:PA2}), and it is equipped with the congruence $e\conv e'$
generated from the same equations (cf section~\ref{ss:KConv}).
The language of formul{\ae} is now the following:
$$\begin{array}{r@{\quad}r@{\quad}l}
  A,B &::=& \Null(e) \quad|\quad \Nat(e)
  \quad|\quad X(e_1,\ldots,e_k) \\
  &|& A\limp B \quad|\quad \forall x\,A \quad|\quad \forall X\,A \\
  &|& A\land B \quad|\quad \exists x\,A \quad|\quad \exists X\,A \\
\end{array}\leqno\begin{array}{l}
  \textbf{Formul{\ae}} \\ \\ \\
\end{array}$$
To the language of formul{\ae} of \PAtwo\ (Fig.~\ref{f:PA2}) we add:
\begin{enumerate}[$\bullet$]
\item A new predicate symbol $\Nat(e)$ to give a type to the
  Peano-style numerals we shall introduce in the language of
  proof-terms.
\item A primitive conjunction $A\land B$ that we shall interpret in the
  intuitionistic realizability model as a type of pairs.
\item Primitive forms of first- and second-order existential
  quantification that will be interpreted in the model as infinitary
  unions (as in~\cite{Oli08}).
\end{enumerate}
In this setting, the units~$\top$ and~$\bot$ are defined with the
shorthands $\top\equiv\exists ZZ$ and $\bot=\forall ZZ$, whereas
numeric quantifications are defined as
$$\begin{array}{r@{~~}c@{~~}l}
  \forallN x\,A(x) &\equiv& \forall x\,(\Nat(x)\limp A(x)) \\[3pt]
  \existsN x\,A(x) &\equiv& \exists x\,(\Nat(x)\land A(x)) \\
\end{array}$$

\subsubsection{The congruence $A\conv A'$}\quad
The congruence $A\conv A'$ over the class of formul{\ae} of HA2 is
defined from the defining equations of the primitive recursive
function symbols of the signature, plus the three equations
$$\begin{array}{c}
  \Null(0)~\conv~\top~\equiv~\exists ZZ\qquad\qquad
  \Null(s(e))~\conv~\bot~\equiv~\forall ZZ \\[6pt]
  (\exists v\,A(v))\limp B~\conv~\forall v\,(A(v)\limp B)
\end{array}\leqno\begin{array}{@{}l}
  \\[3pt]\text{and}\\
\end{array}$$
where $v$ is any first- or second-order variable that does not occur
free in~$B$.
We shall see that the second equation is not only consistent with the
interpretation of existential quantifications as infinitary unions (cf
section~\ref{ss:RealJ}), but that it is also crucial to establish
Prop.~\ref{p:TypingCorrect}.

\subsection{A type system for intuitionistic second-order arithmetic}
\label{ss:TypingHA2}

We introduce an intuitionistic (and more traditional) proof system
based on a judgment of the form $\TYPJ{\Gamma}{t}{A}$, where the
proof-term~$t$ is now formed in the pure $\lambda$-calculus enriched
with the following constants: $\Pair$ (pairing), $\Fst$ (first
projection), $\Snd$ (second projection), $\Zero$ (zero), $\Succ$
(successor) and $\Rec$ (recursor).
In what follows we shall write $\<t;u\>$ for the application
$\Pair\,t\,u$, and denote by~$\Lambda$ the set of all closed
proof-terms.
Typing contexts are simply defined here as finite ordered lists of
declarations of the form $\Gamma\equiv
x_1:A_1,\ldots,x_n:A_n$ where $x_1,\ldots,x_n$ are pairwise distinct
proof-variables.

The class of derivable judgments $\TYPJ{\Gamma}{t}{A}$ is inductively
defined from the rules of inference of Fig.~\ref{f:HA2} (writing
$\forallN x\,A(x)\equiv\forall x\,(\Nat(x)\limp A(x))$).
Note that there is no elimination rule for first- and second-order
primitive existential quantification, since the desired elimination
can be performed using the conversion rule
$\forall v\,(A(v)\limp B)\conv(\exists v\,A(v))\limp B$
(where $v\notin\FV(B)$).

\begin{figure}[p]
  \def\myskip{\vskip 9pt}
  $$\begin{array}{@{}c@{}}
    \hline\hline
    \noalign{\medskip}
    \underline{\textbf{The language of HA2}} \\
    \noalign{\bigskip}
    \begin{array}{l@{\qquad}r@{\quad}r@{\quad}l@{}}
      \textbf{Formul{\ae}}
      & A,B &::=& X(e_1,\ldots,e_k)\quad|\quad
      \Null(e) \quad|\quad \Nat(e) \\
      &&|&  A\limp B \quad|\quad
      \forall x\,A \quad|\quad \forall XA
      \quad|\quad \exists x\,A \quad|\quad \exists XA \\[6pt]
      \textbf{Proof-terms}
      & t,u &::=& x \quad|\quad \Lam{x}{t} \quad|\quad tu
      \quad|\quad \Pair \\
      &&|& \Fst \quad|\quad \Snd \quad|\quad \Zero
      \quad|\quad \Succ \quad|\quad \Rec \\[6pt]
      \textbf{Contexts}
      & \Gamma &::=& \emptyset \quad|\quad \Gamma,x:A \\
    \end{array} \\
    \noalign{\bigskip}
    \underline{\textbf{The congruence $A\conv A'$}} \\
    \noalign{\bigskip}
    \Null(0)\conv\top\qquad
    \Null(s(x))\conv\bot\qquad
    (\exists v\,A)\limp B~\conv~\forall v\,(A\limp B)\quad
    \text{\footnotesize$(v\notin\FV(B))$} \\
    \noalign{\bigskip}
    \underline{\textbf{Abbreviations}} \\
    \noalign{\bigskip}
    \begin{array}{r@{\quad}c@{\quad}l@{\qquad}r@{\quad}c@{\quad}l}
      \top &\equiv& \exists Z\,Z &
      \forallN x\,A(x) &\equiv& \forall x\,(\Nat(x)\limp A(x)) \\
      \bot &\equiv& \forall Z\,Z &
      \existsN x\,A(x) &\equiv& \exists x\,(\Nat(x)\land A(x)) \\
    \end{array} \\
    \noalign{\smallskip}
    e=e'\quad\equiv\quad\forall Z\,(Z(e)\limp Z(e')) \\
    \noalign{\bigskip}
    \underline{\textbf{Typing rules}} \\
    \noalign{\bigskip}
    \infer[\scriptstyle(x:A)\in\Gamma]{\TYPJ{\Gamma}{x}{A}}{}
    \qquad\qquad
    \infer[\scriptstyle A\conv A']{\TYPJ{\Gamma}{t}{A'}}{
      \TYPJ{\Gamma}{t}{A}
    } \\
    \noalign{\myskip}
    \infer{\TYPJ{\Gamma}{\Pair}{A\limp B\limp A\land B}}{} \\
    \noalign{\myskip}
    \infer{\TYPJ{\Gamma}{\Fst}{A\land B\limp A}}{}\qquad\qquad
    \infer{\TYPJ{\Gamma}{\Snd}{A\land B\limp B}}{} \\
    \noalign{\myskip}
    \infer{\TYPJ{\Gamma}{\Zero}{\Nat(0)}}{} \qquad\qquad
    \infer{\TYPJ{\Gamma}{\Succ}
      {\forallN x~\Nat(s(x))}}{} \\
    \noalign{\myskip}
    \infer{\TYPJ{\Gamma}{\Rec}
      {\forall Z\,(Z(0)\limp\forallN y\,(Z(y)\limp Z(s(y)))
        \limp\forallN x\,Z(x))}}{} \\
    \noalign{\myskip}
    \infer{\TYPJ{\Gamma}{\Lam{x}{t}}{A\limp B}}{
      \TYPJ{\Gamma,x:A}{t}{B}
    } \qquad\qquad
    \infer{\TYPJ{\Gamma}{tu}{B}}{
      \TYPJ{\Gamma}{t}{A\limp B} &\quad \TYPJ{\Gamma}{u}{A}
    } \\
    \noalign{\myskip}
    \infer[\scriptstyle x\notin\FV(\Gamma)]
    {\TYPJ{\Gamma}{t}{\forall x\,A}}{
      \TYPJ{\Gamma}{t}{A}
    } \qquad\qquad
    \infer{\TYPJ{\Gamma}{t}{A\{x:=e\}}}{
      \TYPJ{\Gamma}{t}{\forall x\,A}
    } \\
    \noalign{\myskip}
    \infer[\scriptstyle X\notin\FV(\Gamma)]
    {\TYPJ{\Gamma}{t}{\forall XA}}{
      \TYPJ{\Gamma}{t}{A}
    } \qquad\qquad
    \infer{\TYPJ{\Gamma}{t}{A\{X(x_1,\ldots,x_k):=B\}}}{
      \TYPJ{\Gamma}{t}{\forall XA}
    } \\
    \noalign{\myskip}
    \infer{\TYPJ{\Gamma}{t}{\exists x\,A}}{
      \TYPJ{\Gamma}{t}{A\{x:=e\}}
    } \qquad\qquad
    \infer{\TYPJ{\Gamma}{t}{\exists XA}}{
      \TYPJ{\Gamma}{t}{A\{X(x_1,\ldots,x_k):=B\}}
    } \\
    \noalign{\medskip}
    \hline\hline
  \end{array}$$\vspace{-12pt}
  \caption{Intuitionistic second-order arithmetic (HA2)}
  \label{f:HA2}
\end{figure}

The type system of HA2 is expressive enough to provide typable
proof-terms for all the theorems of intuitionistic second-order
arithmetic.
(The specific axioms of arithmetic are treated the same way as
in~PA2.)

\subsection{Weak reduction and inner reduction}

Proof-terms of HA2 are equipped with a binary relation of one-step
\emph{weak reduction} written $t\wred t'$ and defined from the rules
$$\begin{array}{c}
  \infer{(\Lam{x}{t})u\wred t\{x:=u\}}{}\quad~~
  \infer{\vphantom{()}\Rec\,u_0\,u_1\,\Zero\wred u_0}{} \quad~~
  \infer{\Rec\,u_0\,u_1\,(\Succ\,t)\wred
    u_1\,t\,(\Rec\,u_0\,u_1\,t)}{} \\
  \noalign{\medskip}
  \infer{\Fst\,\<t_1;t_2\>\wred t_1}{} \qquad
  \infer{\Snd\,\<t_1;t_2\>\wred t_2}{} \qquad
  \infer{tu\wred t'u}{t\wred t'}\qquad
  \infer{tu\wred tu'}{u\wred u'} \\
\end{array}$$
Note that weak reduction is allowed both in the left- and
right hand-side of applications, but not below $\lambda$-abstraction
(i.e.\ we disable the $\xi$-rule of $\lambda$-calculus).
We write $\wred^*$ the reflexive-transitive closure of one step
weak reduction.

\begin{lem}\label{l:SubstWRed}
  If $t\wred t'$, then
  $t\{x:=u\}\wred t'\{x:=u\}$ (for all terms~$u$).
\end{lem}

\proof
  By induction on the derivation of $t\wred t'$.\qed

Complementarily to the notion of weak reduction, we also define a
relation of \emph{inner reduction} written $t\ired t'$ from the rules:
$$\infer{\Lam{x}{t}\ired\Lam{x}{t'}}{t\wred t'} \qquad
\infer{tu\ired t'u}{t\ired t'} \qquad
\infer{tu\ired tu'}{u\ired u'} \qquad
\infer{\Lam{x}{t}\ired\Lam{x}{t'}}{t\ired t'}$$
The reflexive-transitive closure of the relation of inner reduction is
written $\ired^*$ while its reflexive-symmetric-transitive closure is
written $=_i$.

The union of both relations $\wred$ and $\ired$ is the ordinary
relation of one step reduction, written~$\red$.
By the standard method of parallel reductions we get:
\begin{prop}\label{p:RedCR}
  The relation $\red$ is confluent.
\end{prop}

We now want to deduce from this proposition a result of confluence for
weak reduction modulo inner reductions.
For that, we first need to show that inner reductions can be
postponed, in this sense that any finite sequence of (weak and inner)
reductions can be decomposed into a finite sequence of weak reductions
followed by a finite sequence of inner reductions.
Following Takahashi~\cite{Tak89}, we shall prove this result by
introducing a notion of \emph{parallel inner reduction}, written
$t\jred t'$ and defined from the rules:
$$\infer{t\jred t}{} \qquad
\infer{\Lam{x}{t}\jred\Lam{x}{t'}}{t\wred t'} \qquad
\infer{tu\jred t'u'}{t\jred t' & u\jred u'} \qquad
\infer{\Lam{x}{t}\jred\Lam{x}{t'}}{t\jred t'}$$
From this definition it is clear that
$({\ired})\subseteq({\jred})\subseteq({\ired}^*)$,
so that $({\ired}^*)=({\jred}^*)$.
We first check that parallel inner reduction enjoys the expected
property of substitutivity:
\begin{lem}\label{l:SubstJRed}
  If $t\jred t'$ and $u\jred u'$, then
  $t\{x:=u\}\jred t'\{x:=u'\}$.
\end{lem}

\proof
  By induction on the derivation of $t\jred t'$.\qed

\begin{prop}\label{p:PostInner}
  If $t\jred t'\wred u$, then $t\wred^+u_0\jred u$
  for some term~$u_0$.
\end{prop}

\proof
  By induction on the derivation of $t'\wred u$.\qed

\begin{cor}[Postponement]\label{c:PostInner2}
  If $t\red^*u$, then $t\wred^*u_0\ired^*u$
  for some~$u_0$.
\end{cor}

\proof
  We first show that if $t\jred t'\wred^*u$, then $t\wred^*u_0\jred u$
  for some~$u_0$, by induction on the number of reduction steps
  in $t'\wred^*u$ using Prop.~\ref{p:PostInner}.
  From this we deduce the desired property by induction on the number
  of reduction steps in $t\red^*u$, using the fact that
  $({\jred}^*)=({\ired}^*)$.\qed

From Prop.~\ref{p:RedCR} and Corollary~\ref{c:PostInner2} we
immediately get:
\begin{prop}[Confluence of $\wred$ modulo $=_i$]
  It $t\wred^*t_1$ and $t\wred^*t_2$, then there are terms~$t'_1$
  and~$t'_2$ such that $t_1\wred^*t'_1$, $t_2\wred^*t'_2$ and
  $t'_1=_it'_2$.
\end{prop}

\subsection{The intuitionistic realizability model}
\label{ss:RealJ}

We now build a simple realizability model for the type system defined
above, in which formul{\ae} are interpreted as \emph{saturated sets}
of terms, that is, as sets of closed proof-terms $S\subseteq\Lambda$
such that both conditions $t\wred t'$ and $t'\in S$ imply $t\in S$.
The set of all saturated sets is written~$\SAT$.

Here, a \emph{valuation} is a function~$\rho$ whose domain is a finite
set of (first- and second-order) variables, such that:
\begin{enumerate}[$\bullet$]
\item $\rho(x)\in\N$ for every first-order variable $x\in\dom(\rho)$;
\item $\rho(X):\N^k\to\SAT$
  for every $k$-ary second-order variable $X\in\dom(\rho)$.
\end{enumerate}
Parametric expressions, formul{\ae} and contexts are defined as
before.
Every closed parametric formula~$A[\rho]$ is interpreted as a
saturated set $\Int{A[\rho]}\in\SAT$ that is defined by the expected
equations
$$\begin{array}{rcl@{\hskip-38mm}rcl}
  \Int{X(e_1,\ldots,e_k)[\rho]} &=&
  \rho(X)(\Val(e_1[\rho]),\ldots,\Val(e_k[\rho])) \\
  \noalign{\medskip}
  \Int{\Null(e)[\rho]} &=& \begin{cases}
    \Lambda &\text{if}~\Val(e[\rho])=0 \\
    \varnothing &\text{if}~\Val(e[\rho])\neq 0 \\
  \end{cases} \\
  \noalign{\medskip}
  \Int{\Nat(e)[\rho]} &=& \{t\in\Lambda~:~
  t\wred^*\Succ^n\Zero,~\text{where}~n=\Val(e[\rho])\} \\
  \noalign{\medskip}
  \Int{(A\limp B)[\rho]} &=&
  \{t\in\Lambda~:~\forall u\,{\in}\,\Int{A[\rho]}~~
  tu\in\Int{B[\rho]}\} \\
  \noalign{\medskip}
  \Int{(A\land B)[\rho]} &=&
  \{t\in\Lambda~:~
  \exists t_1\,{\in}\,\Int{A[\rho]}~
  \exists t_2\,{\in}\,\Int{B[\rho]}~~t\wred^*\<t_1;t_2\>\} \\
  \noalign{\bigskip}
  \Int{(\forall x\,A)[\rho]} &=&\ds
  \bigcap_{n\in\N}\Int{A[\rho;x\gets n]} &
  \Int{(\forall XA)[\rho]} &=&\!\!\!\ds
  \bigcap_{\!\!\!\!\!F:\N^k\to\SAT\!\!\!\!\!}\Int{A[\rho;X\gets F]} \\
  \noalign{\medskip}
  \Int{(\exists x\,A)[\rho]} &=&\ds
  \bigcup_{n\in\N}\Int{A[\rho;x\gets n]} &
  \Int{(\exists XA)[\rho]} &=&\!\!\!\ds
  \bigcup_{\!\!\!\!\!F:\N^k\to\SAT\!\!\!\!\!}\Int{A[\rho;X\gets F]} \\
\end{array}$$
In what follows, we shall write $t\realJ A[\rho]$ for
$t\in\Int{A[\rho]}$.

\begin{lem}\label{l:JConvFormSound}
  If~$A$ and~$A'$ are two formul{\ae} of HA2 such that
  $A\conv A'$, then for all valuations~$\rho$ closing~$A$ and~$A'$
  we have $\Int{A[\rho]}=\Int{A'[\rho]}$.
\end{lem}

\subsection{Adequacy}
\label{ss:AdequacyJ}

Given a substitution $\sigma$ and a closed parametric context
$\Gamma[\rho]$, we write $\sigma\realJ\Gamma[\rho]$ when
$\dom(\Gamma)\subseteq\dom(\sigma)$ and
$\sigma(x)\realJ A[\rho]$ for all $(x:A)\in\Gamma$.
We say that:
\begin{enumerate}[$\bullet$]
\item A judgment $\TYPJ{\Gamma}{t}{A}$ is \emph{sound} when for
  all valuations $\rho$ and for all substitutions $\sigma$ such
  that $\sigma\realJ\Gamma[\rho]$, we have
  $t[\sigma]\realJ A[\rho]$.
\item An inference rule $\frac{P_1\cdots P_n}{C}$
  (where~$P_1,\ldots,P_n$ and $C$ are typing judgments) is
  \emph{sound} when the soundness of its premises $P_1,\ldots,P_n$ (in
  the above sense) implies the soundness of its conclusion~$C$.
\end{enumerate}

\begin{prop}[Adequacy]
  The typing rules of Fig.~\ref{f:HA2} are sound.
\end{prop}

From this result, we immediately get:
\begin{prop}[Witness property]\label{p:JWitness}
  If $\TYPJ{}{t}{\existsN x\,\Null(f(x))}$,
  then there are a number $n\in\N$ and a closed term~$u$ such that
  $f(n)=0$ and $t\wred^*\<\Succ^n\Zero;u\>$.
\end{prop}

\proof
  From the definition of the denotation of the formula
  $\existsN x\,\Null(f(x))\equiv\exists x\,(\Nat(x)\land\Null(f(x)))$
  in the realizability model, we know that there are a number $n\in\N$
  and a term $u\in\Lambda$ such that $t\wred^*\<\Succ^n\Zero;u\>$ and
  $u\in\Int{\Null(f(n))}$.
  Which means that the denotation $\Int{\Null(f(n))}$ is inhabited,
  so that $f(n)=0$ (by definition of the interpretation of the
  predicate~$\Null$).\qed

\section{The negative translation}
\label{s:NegTrans}

\subsection{Translating formul{\ae}}
\label{ss:TransForm}

We now define a negative translation of the formul{\ae} of~$\PAext$
(Fig.~\ref{f:PA2} and~\ref{f:PA2ext}) into formul{\ae} of~HA2
(Fig.~\ref{f:HA2}).
We do not consider the usual double negation translation, but
Streicher and Oliva's negative translation~\cite{Oli08}, that is
designed to mimic Krivine's realizability in intuitionistic logic.
Technically, this translation is parameterized by a fixed formula~$R$
(of HA2) that is intended to represent the pole~$\Bot$.
In what follows, we write $\notR A$ as a shorthand for $A\limp R$.

Every formula~$A$ of~$\PAext$ is translated as two formul{\ae} of HA2,
written~$A^{\nn}$ and~$A^{\bot}$.
Intuitively, the intuitionistic formula $A^{\bot}$ represents the type
of stacks facing a classical proof of~$A$; it is mainly built using
the connective $\land$ (representing the operation of consing)
and from the two primitive forms of existential quantifications in HA2
(corresponding to universal quantification from the point of view of
stacks).
The intuitionistic formula $A^{\nn}$---that represents the type of
classical proofs of~$A$---is uniformly defined by
$A^{\nn}\equiv\notR A^{\bot}$.
Formally:
\begin{defi}[Definition of the negative translation]
  The formula~$A^{\bot}$ is defined by induction on~$A$ by
  the equations
  $$\begin{array}{r@{~~}c@{~~}l@{\qquad\qquad}r@{~~}c@{~~}l}
    (X(e_1,\ldots,e_k))^{\bot} &\equiv& X(e_1,\ldots,e_k) &
    (\Null(e))^{\bot} &\equiv& \Null(\Neg(e)) \\
    (A\limp B)^{\bot} &\equiv& A^{\nn}\land B^{\bot} &
    (\forall x\,A)^{\bot} &\equiv& \exists x\,A^{\bot} \\
    (\{e\}\limp B)^{\bot} &\equiv& \Nat(e)\land B^{\bot} &
    (\forall X\,A)^{\bot} &\equiv& \exists X\,A^{\bot} \\
  \end{array}$$
  (using the unary function~`$\Neg$' defined in
  section~\ref{ss:LangPA2}), whereas the formula $A^{\nn}$ is
  defined as
  $A^{\nn}~\equiv~\notR A^{\bot}~\equiv~A^{\bot}\limp R$.
\end{defi}

\begin{rem}\label{r:ConvForall}
  Notice that through this translation, we have
  $$(\forall v\,A(v))^{\nn}~\equiv~
  \exists v\,A(v)^{\bot}\limp R~\conv~
  \forall v\,(A(v)^{\bot}\limp R)~\equiv~\forall v\,(A(v)^{\nn})\,,$$
  using the specific commutation rules of HA2.
  These conversions are crucial for the translation of the
  introduction and elimination rules of first- and second-order
  universal quantifications in the proof of
  Prop.~\ref{p:TypingCorrect}.
\end{rem}

We first check that the translations $A\mapsto A^{\bot}$ and
$A\mapsto A^{\nn}$ are substitutive:
\begin{lem}[Substitutivity]\label{l:TransSubst}
  For all arithmetic expressions~$e$ and for all
  formul{\ae}~$A$ and~$B$ of~$\PAext$:
  \begin{enumerate}[\em(1)]
  \item$(A\{x:=e\})^{\bot}\equiv A^{\bot}\{x:=e\}$
  \item$(A\{x:=e\})^{\nn}\equiv A^{\nn}\{x:=e\}$
  \item$(A\{X(x_1,\ldots,x_k):=B\})^{\bot}\equiv
    A^{\bot}\{X(x_1,\ldots,x_k):=B^{\bot}\}$
  \item$(A\{X(x_1,\ldots,x_k):=B\})^{\nn}\equiv
    A^{\nn}\{X(x_1,\ldots,x_k):=B^{\bot}\}$
  \end{enumerate}
\end{lem}

\proof
  Item~1 is proved by induction on~$A$, and item~2 immediately follows
  from item~1 (since $A^{\nn}\equiv A^{\bot}\limp R$).
  The same for item~3 and item~4.\qed

It is a simple exercise to check that:
\begin{lem}\label{l:ConvSound}
  If $A\conv A'$ ($\PAext$),
  then $A^{\bot}\conv{A'}^{\bot}$ and $A^{\nn}\conv{A'}^{\nn}$ (HA2).
\end{lem}

\proof
  This is obvious for the defining equations of function symbols (that
  are the same in both systems) since the translation does not affect
  arithmetic expressions.
  We only have to check that
  $$(\Null(s(e)))^{\bot}~\equiv~\Null(\Neg(s(e)))~\conv~
  \Null(0)~\conv~\exists ZZ~\equiv~(\forall ZZ)^{\bot}~\equiv
  ~(\bot)^{\bot}\,.$$
  The rest of the proof proceeds by a straightforward induction.\qed

We finally extend the translation $A\mapsto A^{\nn}$ to a translation
$\Gamma\mapsto\Gamma^{\nn}$ that transforms any context~$\Gamma$ of
$\PAext$ into a context~$\Gamma^{\nn}$ of HA2.
This translation is defined by induction on the length of~$\Gamma$ as
follows:
$$\begin{array}{r@{\quad}c@{\quad}l}
  \emptyset^{\nn} &\equiv& \emptyset \\
  (\Gamma,x:A)^{\nn} &\equiv& \Gamma^{\nn},x:A^{\nn} \\
  (\Gamma,x:\{e\})^{\nn} &\equiv& \Gamma^{\nn},x:\Nat(e)\,.
\end{array}$$

\subsection{CPS-translating terms and stacks}
\label{ss:TransProof}

To define the translation of proof-terms, we introduce the convenient
shorthand
$$\LetP{x}{y}{u}{t}~~\equiv~~(\Lam{xy}{t})\,(\Fst\,u)\,(\Snd\,u)
\eqno(\text{`destructing $\Let$'})$$
We first define a translation $t\mapsto t^*$ from proof-terms of
$\PAext$ (Fig.~\ref{f:PA2} and~\ref{f:PA2ext}) into proof-terms terms
of HA2 (Fig.~\ref{f:HA2}).
We will later extend this translation to continuation constants
$k_{\pi}$ and stacks.
Formally:
\begin{defi}[Translation of proof-terms]\label{d:TransProofs}
  We associate to every proof-term of~$\PAext$ a proof-term~$t^*$
  of~HA2 that is inductively defined by:
  $$\begin{array}{@{}r@{\quad}c@{\quad}l@{}}
    x^* &\equiv& x \\
    (tu)^* &\equiv& \Lam{k}{t^*\,\<u^*;k\>} \\[3pt]
    (\Lam{x}{t})^* &\equiv&
    \Lam{k}{\LetP{x}{k'}{k}{t^*\,k'}} \\[3pt]
    (\cc)^* &\equiv& \Lam{k}
    {\LetP{x}{k'}{k}{x\,\<(\Lam{k''}{\LetP{y}{\_}{k''}{y\,k'}});~k'\>}}
    \\[3pt]
    (\widehat{n})^* &\equiv& \Succ^n\,\Zero \\[3pt]
    (\Succ)^* &\equiv& \Lam{k}{}\LetP{x}{k'}{k}{} \\
    &&\hphantom{\Lam{k}{}}\LetP{y}{k''}{k'}{y\,\<\Succ\,x;~k''\>} \\[3pt]
    (\Rec)^* &\equiv& \Lam{k}{}\LetP{z_0}{k'}{k}{} \\
    &&\hphantom{\Lam{k}{}}\LetP{z_1}{k''}{k'}{} \\
    &&\hphantom{\Lam{k}{}}\LetP{x}{k'''}{k''}{}
    \Rec\,z_0\,(\Lam{x'yk_0}{z_1\,\<x';\<\Lam{k_1}{y\,k_1};k_0\>\>})
    \,x\,k''' \\
  \end{array}$$
\end{defi}
Notice how the destructing $\Let$ is used to mimic the destruction of
the stack represented by the continuation variable~$k$.
Also note that the translation of the constant~$\widehat{n}$ does not
start with a continuation abstraction $\Lam{k}{.\,.}$, which reflects
the fact that this construct is not intended to appear in head
position.

We can now prove the following:
\begin{prop}[Correctness w.r.t.\ typing]%
  \label{p:TypingCorrect}
  If $\TYPK{\Gamma}{t}{A}$ in $\PAext$, then
  $\TYPJ{\Gamma^{\nn}}{t^*}{A^{\nn}}$ (in the sense of
  Fig.~\ref{f:HA2}).
\end{prop}

\proof
By induction on the derivation, distinguishing cases according to the
last applied rule.
We first treat the cases of the rules of Fig.~\ref{f:PA2}.
\begin{enumerate}[$\bullet$]
\item Axiom.\d Immediate, since $x^*\equiv x$.
\item Conversion.\ Immediately follows from
  Lemma~\ref{l:ConvSound}.
\item $\limp$-intro.\
  The desired judgment comes from the following derivation:
  \begin{footmath}
    \infer{\TYPJ{\Gamma^{\nn}}
      {\underbrace{\Lam{k}{\LetP{x}{k'}{k}{t^*\,k'}}}
        _{\textstyle(\Lam{x}{t})^*}}
      {(A\limp B)^{\nn}}}{
      \infer{\TYPJ{\Gamma^{\nn},~k:A^{\nn}\land B^{\bot}}
        {\LetP{x}{k'}{k}{t^*\,k'}}{R}}{
        \infer{\TYPJ{\Gamma^{\nn},~k:A^{\nn}\land B^{\bot},~
            x:A^{\nn},~k':B^{\bot}}{t^*\,k'}{R}}{
          \infer={\TYPJ{\Gamma^{\nn},~k:A^{\nn}\land B^{\bot},~
              x:A^{\nn},~k':B^{\bot}}{t^*}{B^{\bot}\limp R}}{
            \infer*[(\text{IH})]
            {\TYPJ{(\Gamma,~x:A)^{\nn}}{t^*}{B^{\nn}}}{}
          }
        }
      }
    }
  \end{footmath}%
  (In the derivation above, we omit obvious branches and indicate
  uses of the admissible rule of weakening with a double bar.)
\item $\limp$-elim.\
  The desired judgment comes from the following derivation:
  \begin{footmath}
    \infer{\TYPJ{\Gamma^{\nn}}
      {\underbrace{\Lam{k}{t^*\,\<u^*;k\>}}_{\textstyle(tu)^*}}{B^{\nn}}}{
      \infer{\TYPJ{\Gamma^{\nn},~k:B^{\bot}}{t^*\,\<u^*;k\>}{R}}{
        \infer={\TYPJ{\Gamma^{\nn},~k:B^{\bot}}{t^*}
          {A^{\nn}\land B^{\bot}\limp R}}{
          \infer*[(\text{IH})]
          {\TYPJ{\Gamma^{\nn}}{t^*}{(A\limp B)^{\nn}}}{}
        } &
        \infer{\TYPJ{\Gamma^{\nn},~k:B^{\bot}}{\<u^*;k\>}
          {A^{\nn}\land B^{\bot}}}{
          \infer={\TYPJ{\Gamma^{\nn},~k:B^{\bot}}{u^*}{A^{\nn}}}{
            \infer*[(\text{IH})]{\TYPJ{\Gamma^{\nn}}{u^*}{A^{\nn}}}{}
          }
        }
      }
    }
  \end{footmath}
\item $\forall$-intro (1st order).\
  The desired judgment comes from the derivation
  \begin{footmath}
    \infer[(\text{Remark~\ref{r:ConvForall}})]
    {\TYPJ{\Gamma^{\nn}}{t^*}{(\forall x\,A)^{\nn}}}{
      \infer{\TYPJ{\Gamma^{\nn}}{t^*}{\forall x\,(A^{\nn})}}{
        \infer*[(\text{IH})]{\TYPJ{\Gamma^{\nn}}{t^*}{A^{\nn}}}{}
      }
    }
  \end{footmath}
\item $\forall$-elim (1st order).\
  The desired judgment comes from the derivation
  \begin{footmath}
    \infer[(\text{Lemma~\ref{l:TransSubst}})]
    {\TYPJ{\Gamma^{\nn}}{t^*}{(A\{x:=e\})^{\nn}}}{
      \infer{\TYPJ{\Gamma^{\nn}}{t^*}{A^{\nn}\{x:=e\}}}{
        \infer[(\text{Remark~\ref{r:ConvForall}})]
        {\TYPJ{\Gamma^{\nn}}{t^*}{\forall x\,(A^{\nn})}}{
          \infer*[(\text{IH})]
          {\TYPJ{\Gamma^{\nn}}{t^*}{(\forall x\,A)^{\nn}}}{}
        }
      }
    }
  \end{footmath}
\item $\forall$-intro (2nd order).\
  The desired judgment comes from the derivation
  \begin{footmath}
    \infer[(\text{Remark~\ref{r:ConvForall}})]
    {\TYPJ{\Gamma^{\nn}}{t^*}{(\forall X\,A)^{\nn}}}{
      \infer{\TYPJ{\Gamma^{\nn}}{t^*}{\forall X\,(A^{\nn})}}{
        \infer*[(\text{IH})]{\TYPJ{\Gamma^{\nn}}{t^*}{A^{\nn}}}{}
      }
    }
  \end{footmath}
\item $\forall$-elim (2nd order).\
  The desired judgment comes from the derivation
  \begin{footmath}
    \infer[(\text{Lemma~\ref{l:TransSubst}})]
    {\TYPJ{\Gamma^{\nn}}{t^*}{(AX(x_1,\ldots,x_k):=B\})^{\nn}}}{
      \infer{\TYPJ{\Gamma^{\nn}}{t^*}
        {A^\nn\{X(x_1,\ldots,x_k):=B^{\bot}\}}}{
        \infer[(\text{Remark~\ref{r:ConvForall}})]
        {\TYPJ{\Gamma^{\nn}}{t^*}{\forall X\,(A^{\nn})}}{
          \infer*[(\text{IH})]
          {\TYPJ{\Gamma^{\nn}}{t^*}{(\forall X\,A)^{\nn}}}{}
        }
      }
    }
  \end{footmath}
\item Peirce's law.\
  Let us use the shorthand
  $u_{k'}\equiv\Lam{k''}{\LetP{y}{\_}{k''}{y\,k'}}$.
  The desired judgment comes from the derivation
  \begin{footmath}
    \infer={\TYPJ{\Gamma^{\nn}}
      {\underbrace{\Lam{k}{\LetP{x}{k'}{k}{x\,\<u_{k'};~k'\>}}}
        _{\textstyle\cc^*}}
      {(((A\limp B)\limp A)\limp A)^{\nn}}}{
      \infer{\TYPJ{}{\Lam{k}{\LetP{x}{k'}{k}{x\,\<u_{k'};~k'\>}}}
        {(((A\limp B)\limp A)\limp A)^{\nn}}}{
        \infer{\TYPJ{k:((A\limp B)\limp A)^{\nn}\land A^{\bot}}
          {\LetP{x}{k'}{k}{x\,\<u_{k'};~k'\>}}{R}}{
          \infer={\TYPJ{k:((A\limp B)\limp A)^{\nn}\land A^{\bot},~
            x:((A\limp B)\limp A)^{\nn},~k':A^{\bot}}
            {x\,\<u_{k'};~k'\>}{R}}{
            \infer{\TYPJ{x:(A\limp B)^{\nn}\land A^{\bot}\limp R,~
                k':A^{\bot}}{x\,\<u_{k'};~k'\>}{R}}{
              \infer{\TYPJ{x:(A\limp B)^{\nn}\land A^{\bot}\limp R,~
                  k':A^{\bot}}{\<u_{k'};~k'\>}
                {(A\limp B)^{\nn}\land A^{\bot}}}{
                \infer={\TYPJ{x:(A\limp B)^{\nn}\land A^{\bot}\limp R,~
                    k':A^{\bot}}{u_{k'}}{(A\limp B)^{\nn}}}{
                  \infer{\TYPJ{k':A^{\bot}}
                    {\Lam{k''}{\LetP{y}{\_}{k''}{y\,k'}}}
                    {(A\limp B)^{\nn}}}{
                    \infer={\TYPJ{k':A^{\bot},~k'':A^{\nn}\land B^{\bot}}
                      {\LetP{y}{\_}{k''}{y\,k'}}{R}}{
                      \infer{\TYPJ{k':A^{\bot},~
                          k'':A^{\nn}\land B^{\bot},~y:A^{\nn}}
                        {y\,k'}{R}}{}
                    }
                  }
                }
              }
            }
          }
        }
      }
    }
  \end{footmath}
\end{enumerate}
Let us now treat the rules of Fig.~\ref{f:PA2ext}.
\begin{enumerate}[$\bullet$]
\item $\{\_\}\limp$-intro.\
  The desired judgment comes from the derivation:
  \begin{footmath}
    \infer{\TYPJ{\Gamma^{\nn}}
      {\underbrace{\Lam{k}{\LetP{x}{k'}{k}{t^*\,k'}}}
        _{\textstyle(\Lam{x}{t})^*}}
      {(\{e\}\limp B)^{\nn}}}{
      \infer{\TYPJ{\Gamma^{\nn},~k:\Nat(e)\land B^{\bot}}
        {\LetP{x}{k'}{k}{t^*\,k'}}{R}}{
        \infer{\TYPJ{\Gamma^{\nn},~k:\Nat(e)\land B^{\bot},~
            x:\Nat(e),~k':B^{\bot}}{t^*\,k'}{R}}{
          \infer={\TYPJ{\Gamma^{\nn},~k:\Nat(e)\land B^{\bot},~
              x:\Nat(e),~k':B^{\bot}}{t^*}{B^{\bot}\limp R}}{
            \infer*[(\text{IH})]
            {\TYPJ{(\Gamma,~x:\{e\})^{\nn}}{t^*}{B^{\nn}}}{}
          }
        }
      }
    }
  \end{footmath}
\item $\{\_\}\limp$-elim-1.\
  The desired judgment comes from the derivation:
  \begin{footmath}
    \infer{\TYPJ{\Gamma^{\nn}}
      {\underbrace{\Lam{k}{t^*\,\<x;k\>}}_{\textstyle(t\,x)^*}}{B^{\nn}}}{
      \infer{\TYPJ{\Gamma^{\nn},~k:B^{\bot}}{t^*\,\<x;k\>}{R}}{
        \infer={\TYPJ{\Gamma^{\nn},~k:B^{\bot}}{t^*}
          {\Nat(e)\land B^{\bot}\limp R}}{
          \infer*[(\text{IH})]
          {\TYPJ{\Gamma^{\nn}}{t^*}{(\{e\}\limp B)^{\nn}}}{}
        } &
        \infer{\TYPJ{\Gamma^{\nn},~k:B^{\bot}}{\<x;k\>}
          {\Nat(e)\land B^{\bot}}}{
          \infer%[(x:\{e\})\in\Gamma]
          {\TYPJ{\Gamma^{\nn},~k:B^{\bot}}{x}{\Nat(e)}}{}
        }
      }
    }
  \end{footmath}
\item $\{\_\}\limp$-elim-2.\
  The desired judgment comes from the derivation:
  \begin{footmath}
    \infer{\TYPJ{\Gamma^{\nn}}
      {\underbrace{\Lam{k}{t^*\,\<\Succ^n\Zero;k\>}}
        _{\textstyle(t\,\hat{n})^{*}}}
      {B^{\nn}}}{
      \infer{\TYPJ{\Gamma^{\nn},~k:B^{\bot}}{t^*\,\<\Succ^n\Zero;k\>}{R}}{
        \infer={\TYPJ{\Gamma^{\nn},~k:B^{\bot}}{t^*}
          {\Nat(n)\land B^{\bot}\limp R}}{
          \infer*[(\text{IH})]
          {\TYPJ{\Gamma^{\nn}}{t^*}{(\{n\}\limp B)^{\nn}}}{}
        } &
        \infer{\TYPJ{\Gamma^{\nn},~k:B^{\bot}}{\<\Succ^n\,\Zero;k\>}
          {\Nat(n)\land B^{\bot}\!\!}}{
          \infer{\TYPJ{\Gamma^{\nn},~k:B^{\bot}}{\Succ^n\Zero}{\Nat(n)}}{}
        }
      }
    }
  \end{footmath}
\end{enumerate}
The cases of~$\Succ$ and~$\Rec$ are left to the reader.\qed

\subsubsection{Extending the translation to the full
  $\lambda_c$-calculus}\
We now extend the translation $t\mapsto t^*$ defined on the
proof-terms of $\PAext$ into a full translation of the
$\lambda_c$-calculus.
For that, we close the set of instructions~$\K$ by letting
$$\K~=~\{\cc;\Succ;\Rec;\Stop\}\cup\{\widehat{n}~:~n\in\N\}\,,$$
and we close the relation of evaluation~$\eval$ by defining it as the
union of the rules (\textsc{Grab}), (\textsc{Push}),
(\textsc{Call$/$cc}), (\textsc{Resume}), (\textsc{Succ}),
(\textsc{Rec-0}) and (\textsc{Rec-S}).

Formally, we define by mutual induction on~$t$ and~$\pi$ two
translations $t\mapsto t^*$ (where~$t$ now ranges over all terms of
the $\lambda_c$-calculus) and $\pi\mapsto\pi^*$ by adding to the
equations of Def.~\ref{d:TransProofs} the following:
$$\begin{array}{r@{\quad}c@{\quad}l}
  (\k_{\pi})^* &\equiv&
  \Lam{k}{\LetP{x}{\_}{k}{x\,\pi^*}} \\
  \Stop^{*} &\equiv& \Lam{z}{z} \\
\end{array}\qquad\qquad
\begin{array}{r@{\quad}c@{\quad}l}
  (\diamond)^* &\equiv& 0 \\
  (t\cdot\pi)^* &\equiv& \<t^*;~\pi^*\>$$
\end{array}$$
Stacks are translated here in the obvious way, that is: as finite
lists.
Note that the bottom of the stack~$\diamond$ could be translated by
any closed term as well: it has no evaluation rule, and it is not
involved in the type system of PA2.
On the other hand, we translate~$\Stop$ as the identity term, and
this choice will be important in the analysis of
section~\ref{ss:Compare}.

Finally, processes are translated by letting:
$$(t\star\pi)^*\quad\equiv\quad t^*\,\pi^*\,.$$

\subsection{Simulation of evaluation by weak reduction}

The expected property would be that each evaluation step
$t_1\star\pi_2\eval t_2\star\pi_2$ in~$\lambda_c$ corresponds to
one or several weak reduction steps
$t_1^*\,\pi_1^*\wred^+t_2^*\,\pi_2^*$
through the CPS-translation.
Although this works for almost all the evaluation rules (application,
abstraction, call$/$cc, continuation and successor), the property
does not hold for the rule (\textsc{Rec-s}) so that we need to
refine a little bit more.

\begin{prop}[One step simulation]\label{p:OneStepSimul}
  If $t_1\star\pi_1\eval t_2\star\pi_2$ (one step evaluation
  in~$\lambda_c$), then 
  $t_1^*\,\pi_1^*\wred^+t_2^*\,u$ (weak reduction) for some
  term $u=_i\pi_2^*$.
  Moreover, for all rules but (\textsc{Succ-s}),
  we have $u\equiv\pi_2^*$.
\end{prop}

\proof
  We distinguish cases according to the applied rule.
  The cases of abstraction, application, call$/$cc and continuation
  constants are easy---they do not involve inner conversion
  steps---and standard~\cite{Oli08}, so that we do not treat them
  here.
  Let us consider the evaluation rules dealing with primitive
  numerals.
  \begin{enumerate}[$\bullet$]
  \item Rule (\textsc{Succ})\ We have:
    $$\begin{array}{r@{\quad}l@{\quad}l}
      (\Succ\star\widehat{n}\cdot u\cdot\pi)^*
      &\equiv& \Succ^*\,\<\Succ^n\,\Zero;\<u^*;\pi^*\>\> \\
      &\wred^*& u^*\,\<\Succ^{n+1}\,\Zero;~\pi^*\>
      ~~\equiv~~(u\star\widehat{n+1}\cdot\pi)^* \\
    \end{array}$$
  \item Rule (\textsc{Rec-0})\quad We have:
    $$\begin{array}{r@{\quad}l@{\quad}l}
      (\Rec\star u_0\cdot u_1\cdot\widehat{0}\cdot\pi)^*
      &\equiv& \Rec^*\,\<u_0^*;\<u_1^*;\<\Zero;\pi^*\>\>\> \\
      &\wred^*& \Rec\,u_0^*\,T[u_1^*]\,\Zero\,\pi^* \\
      &\wred^*& u_0^*\,\pi^* ~~\equiv~~ (u_0\star\pi)^*\,, \\
    \end{array}$$
    writing $T[z]\equiv
    \Lam{x'yk_0}{z\,\<x';\<\Lam{k_1}{y\,k_1};k_0\>\>}$.
  \item Rule (\textsc{Rec-s})\
    We have:
    $$\begin{array}{r@{\quad}l@{\quad}l}
      (\Rec\star u_0\cdot u_1\cdot\widehat{n+1}\cdot\pi)^*
      &\equiv& \Rec^*\,\<u_0^*;\<u_1^*;\<\Succ^{n+1}\,\Zero;\pi^*\>\>\>\\
      &\wred^*& \Rec\,u_0^*\,T[u_1^*]\,(\Succ^{n+1}\,\Zero)\,\pi^* \\
      &\wred^*& u_1^*\,\<\Succ^n\,\Zero;
      \<\Lam{k_1}{\Rec\,u_0^*\,T[u_1^*]\,(\Succ^n\,\Zero)\,k_1};~\pi^*\>\>\\
    \end{array}$$
    Moreover:
    $$\begin{array}{l@{\quad}l}
      & u_1^*\,\<\Succ^n\,\Zero;~\<\Lam{k_1}{\Rec\,u_0^*\,T[u_1^*]\,
        (\Succ^n\,\Zero)\,k_1};~\pi^*\>\> \\
      =_i& u_1^*\,\<\Succ^n\,\Zero;~
      \<\Lam{k_1}{(\Rec\,u_0\,u_1\,\widehat{n})^*\,k_1};~\pi^*\>\> \\
      =_i& u_1^*\,\<\Succ^n\,\Zero;~
      \<(\Rec\,u_0\,u_1\,\widehat{n})^*;~\pi^*\>\> ~~\equiv~~
      (u_1\star\widehat{n}\cdot(\Rec\,u_0\,u_1\,\widehat{n})\cdot\pi)^*
    \end{array}$$
    (In the second last line, we mimic $\eta$-reduction with an inner
    reduction step, using the fact that the term
    $(\Rec\,u_0\,u_1\,\widehat{n})^*$ is an abstraction).\qed
  \end{enumerate}

\begin{cor}[Grand simulation]
  If $t_1\star\pi_1\eval^* t_2\star\pi_2$ (evaluation
  in~$\lambda_c$), then $t_1^*\,\pi_1^*\wred^*u$ (weak reduction) for
  some term $u=_it_2^*\,\pi_2^*$.
\end{cor}

\proof
  By induction on the number of evaluation steps, using
  Prop.~\ref{p:OneStepSimul} and Corollary~\ref{c:PostInner2} for the
  induction case.\qed

\subsection{The negative interpretation of classical witness extraction}
\label{ss:Compare}

Let us now reinterpret the classical witness extraction method
described in section~\ref{ss:ExtrSigma01} through the negative
translation defined above.

For that, let us consider a closed classical proof term~$t_0$ such
that
$$\TYPK{}{~t_0~}{~\existsN x\,f(x)=0}$$
(where $\existsN x\,f(x)=0\equiv
\forall Z\,(\forall x\,(\{x\}\limp f(x)=0\limp Z)\limp Z)$),
and let us analyze the behavior of the process
$$p_0~\equiv~t_0\star(\Lam{xy}{y\,(\Stop\,x)})\cdot\diamond$$
that performs witness extraction (by Prop.~\ref{p:ExtrSigma01})
through the negative translation of section~\ref{ss:TransForm} and
the CPS-translation of section~\ref{ss:TransProof}.
(We can end~$p_0$ with the empty stack from the results of
section~\ref{ss:Independence}.)

In the sequel, we write $u\equiv\Lam{xy}{y\,(\Stop\,x)}$.

\subsubsection{Typing the process $(p_0)^*$}\
From Prop.~\ref{p:TypingCorrect} we get
$$\TYPJ{}{~t_0^*~}{~\forall Y\,\Bigl(
  \forall x\,\bigl(\Nat(x)\land(f(x)=0)^{\nn}\land Y\limp R\bigr)
  \land Y\limp R\Bigr)}$$
(writing conjunction right-associative), so that by instantiating~$Y$
with~$\top$:
$$\TYPJ{}{~t_0^*~}{~
  \forall x\,\bigl(\Nat(x)\land(f(x)=0)^{\nn}\land\top\limp R\bigr)
  \land\top\limp R}\,.$$
Let us now fix the pole~$R$ by letting
$R\equiv\existsN x\,\Null(f(x))$.
Since $\Stop^*\equiv\Lam{z}{z}$, we can give it the type
$$\TYPJ{}{~\Stop^*~}{~\forall x\,
  \bigl(\Nat(x)\land\Null(f(x))\limp R\bigr)}\,,$$
which is precisely the introduction rule of numeric existential
quantification.
(Remember that
$R\equiv\existsN x\,\Null(f(x))\equiv
\exists x\,(\Nat(x)\land\Null(f(x)))$.)

Thanks to this, we can typecheck through the CPS-translation all the
constituents of the term~$u$.
We first have
$$\TYPJ{x:\Nat(x)~}{~(\Stop\,x)^*~}{~\Null(f(x))\limp R}\,.$$
Moreover:
$$\TYPJ{y:(f(x)=0)^{\nn}~}{~y~}{~\forall Z\,\Bigl(
  \bigl(Z(f(x))\limp R\bigr)\land Z(0)\limp R\Bigr)}$$
(using the definition of Leibniz equality, the axiom rule and the
conversion rule), so that by instantiating
$Z(x)$ with $\Null(x)$ we get
$$\TYPJ{y:(f(x)=0)^{\nn}~}{~y~}{
  \bigl(\Null(f(x))\limp R\bigr)\land\top\limp R}$$
We thus have
$$\TYPJ{x:\Nat(x),~y:(f(x)=0)^{\nn}~}{(y\,(\Stop\,x))^*~}{
  \top\limp R}$$
and finally:
$$\TYPJ{}{~u^*~}{
  \forall x\,\bigl(\Nat(x)\land(f(x)=0)^{\nn}\land\top\limp R\bigr)}$$
We can now typecheck the term $(u\cdot\diamond)^*$
$$\TYPJ{}{~(u\cdot\diamond)^*~\equiv~\<u^*;0\>~}{~
  \bigl(\Nat(x)\land(f(x)=0)^{\nn}\land\top\limp R\bigr)\land\top}\,,$$
so that $\TYPJ{}{~p_0^*~\equiv~t_0^*\,\<u^*;0\>~}{~R}$.

This shows that the CPS-translation of the process described in
Prop.~\ref{p:ExtrSigma01} is actually an intuitionistic proof (in
HA2) of the formula $R\equiv\existsN x\,\Null(f(x))$.
From Prop.~\ref{p:JWitness}, we thus know that the term $(p_0)^*$
weakly reduces to a pair whose first component is the desired witness.

\medbreak
Through the CPS-translation defined in section~\ref{ss:TransProof},
the extraction method described in section~\ref{ss:ExtrSigma01} thus
amounts to transform a classical proof-term~$t_0$ of the formula
$\existsN x\,f(x)=0$ into an intuitionistic proof-term
$t_0^*\<u^*;0\>$ of the same formula (up to the coding details of
numeric existential quantification and of the predicate expressing the
nullity of its argument).
Here we can see the essential ingredients of Friedman's
transformation:
\begin{enumerate}[$\bullet$]
\item The use of a negative translation to transform a classical
  proof~$t_0$ of a $\Sigma^0_1$-formula into an intuitionistic
  proof~$t_0^*$ of a more complex formula.
\item The choice of the return formula~$R$ (the pole), that is
  precisely defined as the formula we want to prove
  intuitionistically.
\item The key use of the introduction rule of numeric existential
  quantification (here via the term $\Stop^*\equiv\Lam{z}{z}$) to
  return the desired result.
\end{enumerate}

\section{Conclusion}

\subsection{From BHK semantics to Krivine's semantics}

Friedman's extraction method consists to transform a classical proof
of an existential formula into an intuitionistic proof of the same
formula.
Its main drawback is that the underlying CPS-transformation makes the
resulting program much bigger than the originating proof, and
more difficult to understand.

For this reason, much attention has been devoted to the optimization
of the extracted code.
The typical approach is \emph{refined program
  extraction}~\cite{BBS02}, that relies on a clever analysis of
formul{\ae} in order to minimize the insertion of negations during the
translation.
In practice, such an approach gives much better programs than
Friedman's method, typically when considering examples such as the
one we treated in section~\ref{s:Example}.
However, the approach of refined program extraction ultimately remains
intuitionistic, since the extracted program is built and analyzed
according to the traditional Brouwer-Heyting-Kolmogorov (BHK)
semantics, using the tools of intuitionistic realizability (i.e.\ the
mathematical expression of BHK semantics).

In this paper, we have proposed another approach, which is to extract
a program from a classical proof directly, by interpreting classical
reasoning principles with control operators.
The price to pay is that the computational meaning of the extracted
program cannot be analyzed within the traditional BHK semantics
anymore.
For that, we proposed to use Krivine's theory of classical
realizability, that constitutes a genuine alternative to BHK semantics
for classical logic.

\subsubsection{A negative semantics for classical programs}\quad
It has already been pointed out~\cite{Oli08} that Friedman's negative
translation is hard-wired in Krivine's semantics.
Taking the notations of section~\ref{s:NegTrans}, we get the
correspondence:
\begin{center}
  \begin{tabular}{c@{\qquad\qquad}c}
    \textbf{Krivine's semantics} &
    \textbf{Friedman's translation} \\
    The pole~$\Bot$ & The formula~$R$ \\
    Falsity value~$\|A\|$ & Formula $A^{\bot}$ \\
    Truth value~$|A|=\|A\|^{\Bot}$ &
    Formula $A^{\nn}\equiv A^{\bot}\limp R$ \\
    Classical proof-term~$t$ & CPS-translated term~$t^*$ \\
  \end{tabular}
\end{center}
In this paper, we have shown that the correspondence can be lifted up
at the level of the witness extraction methods, in the sense that the
natural extraction method that comes with Krivine's machinery
(section~\ref{ss:ExtrSigma01}) is, up to a CPS-translation, the same
as Friedman's (provided we use Streicher and Oliva's negative
translation instead of the traditional not-not translation).

However, Krivine's semantics is more subtle than the composition of a
negative translation with the standard BHK interpretation, since the
way it is formulated makes the negative translation implicit.
Thanks to this, it is possible to reason about classical programs
directly.
We illustrated this point with the witness extraction procedures
presented in section~\ref{s:Witness} and with the construction by hand
of a universal realizer of the minimum principle in
section~\ref{s:Example}.

It should also be noted that Krivine's semantics is compatible with
the usual deduction rules of intuitionistic logic, as soon as they are
formulated in FA2 style~\cite{Kri93}.
In particular, the typing rules of Fig.~\ref{f:PA2} but the last one
(Peirce's law) are the usual Curry-style typing rules of
intuitionistic minimal logic, formulated the usual way.
Any intuitionistic proof-term that is well typed in FA2 is not only
correct w.r.t.\ the usual intuitionistic realizability semantics of
FA2~\cite{Kri93}, but it is also correct according to Krivine's
semantics.
The latter departs from the traditional BHK semantics only when
classical reasoning is involved.

\subsection{Executing extracted programs}

In our approach, extracted programs are not ordinary $\lambda$-terms,
but $\lambda$-terms with control operators that need to be evaluated
according to a strict call-by-name discipline.
Specific tools are thus required to execute them%
\footnote{We cannot directly use the interpreters and compilers
  dedicated to popular functional programming languages such as LISP,
  Caml, SML or Haskell, since these tools implement either the
  call-by-value discipline or the call-by-need discipline.}.

The main implementation difficulty comes from stacks, whose machine
representation has to be carefully designed in order to avoid
unnecessary duplications when call$/$cc is executed.
Fortunately, control operators have been introduced in programming
languages long before the discovery of their connection with classical
logic, and we can benefit from the many implementation strategies that
have been proposed since.
To illustrate this, we shall discuss two of them.
(For a survey on the different ways to implement control,
see~\cite{CHO99}.)

\subsubsection{Stacks as chained lists}\quad
The simplest way to implement stacks is to represent them
as heap-allocated chained lists of closures.
With this representation, call$/$cc comes for free since it simply
consists to make a copy of the current stack pointer.
The main advantage of this method is that it naturally maximizes the
possibility of sharing large final segments of stacks.
Its main drawback is that each \textsc{Push} operation requires the
allocation of a small block on the heap, which block is subject to
later garbage collection.
Moreover, the resulting fragmentation of the stack may considerably
degrade cache performance.

However, the simplicity of this approach makes it well-suited for a
small interpreter intended to quickly test small examples.
This is this design that is currently used in the \texttt{jivaro}
machine~\cite{Jiv09}.

\subsubsection{The stack$/$heap model}\quad
In the perspective of implementing a real compiler of
$\lambda_c$-programs, a more realistic representation of stacks is
given by the stack$/$heap model~\cite{CHO99}, where the logical stack
is physically split in two parts:
\begin{enumerate}[$\bullet$]
\item A \emph{stack cache}, that consists of a mutable array of
  closures representing the topmost part of the logical stack.
  This stack cache lies in a fixed zone of the memory, and works
  almost as the ordinary system stack.
\item A \emph{far stack}, that represents the rest of the logical
  stack in the heap as a chained list of non mutable blocks
  containing stack chunks.
  As for all the other heap-allocated blocks, the stack chunks of the
  far stack may be shared and are subject to garbage collection.
\end{enumerate}\medskip

\begin{figure}[htp]
  \begin{center}
    \begin{tabular}{c}
      \hline\hline
      \noalign{\medskip}
      \ifpdf\includegraphics[width=100mm]{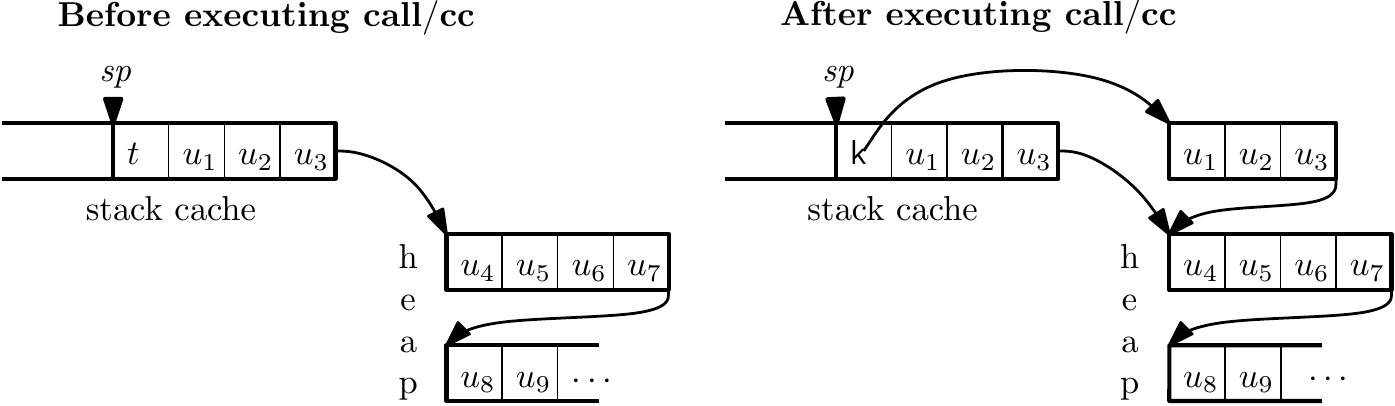}
      \else\includegraphics[width=100mm]{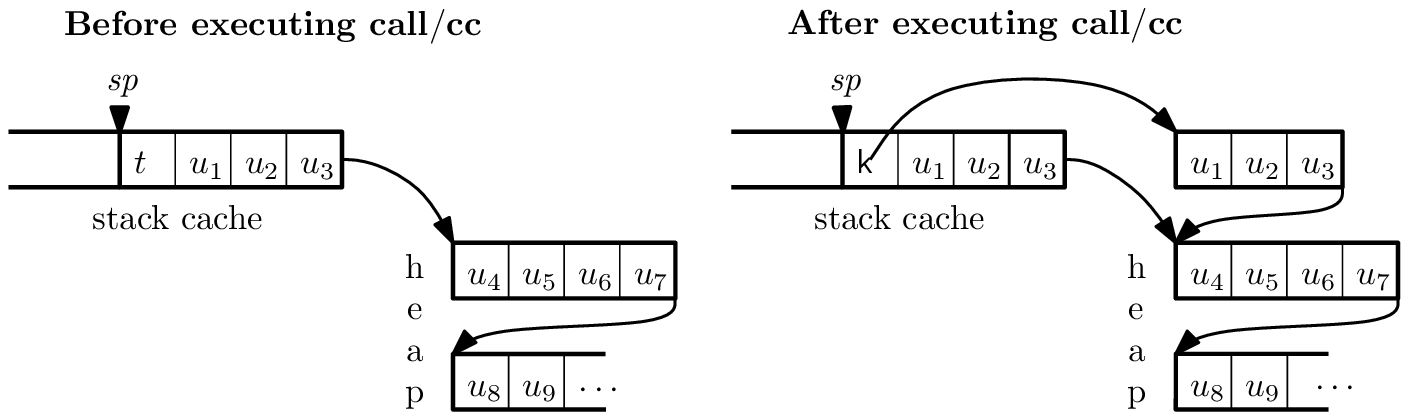}\fi\\
      \noalign{\medskip}
      \hline\hline
    \end{tabular}
  \end{center}\vspace{-12pt}
  \caption{Execution of call$/$cc in the stack$/$heap model}
  \label{f:Stack}
\end{figure}

\noindent The underlying idea of the stack$/$heap approach is that during
execution, almost all the operations on the logical stack take place
in the stack cache, the manipulation of the far stack being
exceptional.
Pushing an argument onto the stack proceeds in the stack cache as
usual, as well as grabbing the topmost element.
The difference is that in the latter case, the cache may underflow, in
which case we need to refill the cache by copying the contents of the
first block of the far stack.
(After copying, the far stack pointer should point to the next
block).
With this approach, call$/$cc only needs to make a copy of the stack
cache into a newly heap-allocated block.
The pointer to this newly allocated block then becomes the
corresponding continuation constant as illustrated in
Fig.~\ref{f:Stack}.
Restoring a formerly saved stack thus consists to clear the stack
cache and to let the far stack pointer point to the first continuation
block.
(The stack cache will then be automatically refilled with the next
\textsc{Grab} operation.)

The interest of this approach is that call$/$cc only needs to copy the
part of the stack that has been used since the last execution of a
control operator.
The authors of~\cite{CHO99} consider that this approach is a
\emph{zero overhead} approach, in the sense that it adds a negligible
overhead to the most frequent operations \textsc{Push} and
\textsc{Grab}, compared with the traditional single-chunk stack
model.

\subsubsection{On the frequency of control in classical proofs}\quad
The above discussion about implementation issues raises a strong
argument in favor of using $\lambda_c$ for interpreting classical
proofs.
A quick look at the classical proofs of well-known theorems shows that
classical reasoning is definitely not used with the same frequency as
intuitionistic reasoning.
Purely intuitionistic reasoning (introduction/elimination of
connectives and quantifiers, induction\dots) appears everywhere,
whereas classical reasoning principles are only used at some few
strategic places.
In some sense, one can consider that actual mathematics are more
quasi-intuitionistic than really classical.
The execution trace (Fig.~\ref{f:ExecTrace} p.~\pageref{f:ExecTrace})
of the example of section~\ref{s:Example} makes the comparison more
dramatic, since it shows that intuitionistic operations are executed
several hundreds of times while classical operations are executed a
dozen of times (call$/$cc being invoked only once).

These figures suggest that a good execution policy for classical
proofs should concentrate all the execution overhead induced by the
presence of classical reasoning to the classical operations themselves
(that are the less frequent ones) while keeping ordinary
intuitionistic operations (the most frequent ones) as fast as
possible.
But this is precisely what the $\lambda_c$-calculus does, especially
when executed in the stack$/$heap model described above.
On the other hand, using a negative translation---even
optimized---adds a non negligible execution overhead to all the
intuitionistic operations of the proof, just to remove the need of
introducing specific operators for classical reasoning.

\subsection{Classical extraction in Coq}

The ideas presented in this paper have been implemented in a classical
extraction module for Coq developed by the author~\cite{Kex09}.
On the theoretical side, this implementation is based on an extension
of Krivine's classical realizability model to the calculus of
constructions with universes~\cite{Miq07}.
This module permits the extraction of a $\lambda_c$-term from any
classical proof formalized in Coq---provided classical logic is only
allowed in the impredicative sort~\texttt{Prop}.
It also proposes witness extraction facilities based on the techniques
presented in section~\ref{s:Witness}.

This module automatically performs several optimizations in the
extracted code.
For instance, Coq unary numerals (as well as the corresponding
arithmetic operators) are automatically translated into the
primitive numerals discussed in section~\ref{s:PrimInt}.
Similar optimizations are introduced to change the representation of
other inductively defined data-types such as ordering.
Moreover, the extractor proposes predefined optimized realizers for
many theorems of Coq's standard library, following the spirit of what
we did in section~\ref{s:Example} with the minimum principle.
In this way, the user can formalize classical proofs using the tools
provided by Coq's standard library while benefiting from many
optimizations that are allowed by the theory of classical
realizability.

However, these hand-made optimized realizers are provided only for a
little fragment of Coq's standard library, and there is currently no
general mechanism to generate them on the fly.
Future work includes the design of a general theory for realizer
optimization in the framework of classical realizability, following
the spirit of refined program extraction~\cite{BBS02}.

% Biblio
\nocite{Bar84,Gir89,Kri01,Kri03,Kri05,Miq07}
\bibliographystyle{plain}
\bibliography{paper}

\end{document}